\documentclass[12pt,a4paper,final]{article}
\usepackage{booktabs,tabularx,amsmath,array}
\newcolumntype{Y}{>{\centering\arraybackslash}X}

\usepackage{mathtools}

\usepackage{subfiles}
\usepackage[right]{showlabels} 

\usepackage{xcolor}

\usepackage{siunitx}
\usepackage{pdflscape} 
\usepackage{graphicx}
\usepackage{subfig}      
\usepackage[font=small]{caption}

\newcommand{\fourcolw}{0.4\textwidth} 

\usepackage{lmodern} 
\usepackage[utf8]{inputenc}
\usepackage[T1]{fontenc}

\usepackage[english]{babel}

\usepackage{tabularx}
\usepackage{graphicx} 
\usepackage{adjustbox}

\usepackage{caption}
\usepackage{booktabs} 
\usepackage{siunitx}

\usepackage{subfig}

\usepackage{csquotes}
\usepackage[
    giveninits=true, 
    backend=biber, 
    bibstyle=authoryear, 
    citestyle=authoryear-comp, 
    dashed=false, 
    doi=true, 
    maxbibnames=99, 
    maxcitenames=99, 
    uniquename=false 
]{biblatex} 
\DeclareFieldFormat[article]{volume}{\bibstring{jourvol}\addnbspace #1} 
\DeclareFieldFormat[article]{number}{\bibstring{number}\addnbspace #1} 
\renewbibmacro*{volume+number+eid}{
  \printfield{volume}%
  \setunit{\addcomma\space}%
  \printfield{number}%
  \setunit{\addcomma\space}%
  \printfield{eid}
}
\renewbibmacro{in:}{} 

\usepackage{algpseudocodex}

\usepackage{lscape}
\usepackage{enumitem}

\usepackage{amsmath} 
\usepackage{amssymb} 
\usepackage{amsthm} 
\usepackage{mathrsfs}
\usepackage{fullpage} 

\usepackage{float}

\allowdisplaybreaks

\newtheorem{assumption}{Assumption}[section]
\newtheorem{theorem}{Theorem}[section]

\newtheorem{lemma}{Lemma}[section]
\newtheoremstyle{boldremark}
    {\dimexpr\topsep/2\relax} 
    {\dimexpr\topsep/2\relax} 
    {}          
    {}          
    {\bfseries} 
    {.}         
    {.5em}      
    {}          
\theoremstyle{boldremark}

\graphicspath{ {./images/} }
\usepackage{dcolumn}

\usepackage{array}
\newcolumntype{E}{D{.}{.}{2,3}}
\newcolumntype{C}{>{$}c<{$}} 
\newcolumntype{R}{>{$}r<{$}} 
\usepackage{multirow} 

\usepackage{authblk}

\usepackage[margin=2.5cm]{geometry} 

\usepackage[nodayofweek]{datetime} 

\usepackage[UKenglish]{isodate} 

\usepackage[colorlinks=false,urlcolor=blue,citecolor=blue,linkcolor=blue,bookmarks,bookmarksopen=true,bookmarksnumbered=true]{hyperref}

\DeclareMathOperator*{\argmin}{argmin}
\DeclareMathOperator{\Var}{Var} 
\DeclareMathOperator{\Cov}{Cov}
\usepackage{xcolor}

\definecolor{orange}{RGB}{255,165,0}

\definecolor{darkgreen}{RGB}{0,100,0}

\newcommand{\LM}{\textit{LM}}
\newcommand{\IC}{\textit{IC}}

\title{Testing the order of fractional integration when smooth deterministic trends are possibly present\thanks{We are grateful to Fabrizio Iacone for helpful suggestions on an earlier version. }}
\author[1]{Mustafa R. K{\i}l{\i}n\c{c}}
\affil[1]{WHU -- Otto Beisheim School of Management,\linebreak Chair of Econometrics and Statistics, Vallendar, Germany\vspace{1ex}}
\author[1,2]{Michael Massmann\footnote{corresponding author: \href{mailto:michael.massmann@whu.edu}{\texttt{michael.massmann@whu.edu}}
}$^,$}
\affil[2]{Vrije Universiteit, Department of Econometrics and Data Science, Amsterdam, The Netherlands}

\begin{document}

\maketitle

\begin{abstract}
This paper introduces a test for fractional integration in a model that possibly contains smooth deterministic trends. We model the trend component using a Chebyshev polynomial and specify the short-run dynamics semi-parametrically, accommodating a broad class of possibly nonlinear processes, including those with conditional heteroskedasticity. We use a local Whittle approach for constructing a Lagrange multiplier test statistic and for constructing a frequency-domain information criterion for the selection of the order of the Chebyshev polynomial. We show that widely used time-domain information criteria are generally inconsistent for the true order, whereas our frequency-domain criterion remains robust under both short- and long-memory behaviour. Monte Carlo simulations and an empirical application to the UK Great Ratios support our theoretical findings.
  

  
\end{abstract}

\clearpage

\section{Introduction}

The intricate interplay of structural change and long memory is a well-documented phenomenon. Early research by \textcite{diebold2001long}, \textcite{granger2004occasional} and \textcite{sibbertsen2004long} illustrates how a short memory process with (random) regime changes can be mistaken to be a long memory process. In essence, tests for fractional order of integration based on models that erroneously do not account for structural change are generally uninformative. \textcite{iacone2022semiparametric} illustrate this by explicitly modelling structural change by step-indicators and showing that, for instance, the Lagrange multiplier test discussed in \textcite{lobato1998nonparametric} is uninformative under structural breaks because its test statistic diverges under the null. This limitation motivates \textcite{iacone2022semiparametric} to extend the model of \textcite{lobato1998nonparametric}  by using the procedure suggested by \textcite{lavielle2000least} to model and locate the structural breaks. They use standard information criteria for determining the number of breaks and a local Whittle approach for avoiding the need to explicitly model any short-range dependence in the data. 

Another possibility for modelling structural change is via smooth trends instead of step-indicators, see e.g.\ \textcite{harvey2010testing}. \textcite{cuestas2016testing} use a Chebyshev polynomial to model smooth trends when assessing the fractional integration order. \textcite{bierens1997testing} and \textcite{bierens2010time} had shown that these trigonometric functions are particularly effective in approximating any square integrable and differentiable function of time arbitrarily closely. Additionally, trigonometric functions are useful for capturing regular fluctuations in time series data, referred to as cyclical trends by \textcite{anderson1994statistical} and discussed further by \textcite{perron2020trigonometric}. They are also employed to represent seasonal, trend, and irregular components of time series, as described by \textcite{harvey1993time}. \textcite{cuestas2016testing} employ a general-to-specific testing procedure for determining the order of the Chebyshev polynomial and model the short-run dynamics fully parametrically. Yet the difficulty to disentangle smooth trends and long memory is also well-documented, cf., for instance, \textcite{Bhattacharya_Gupta_Waymire_1983}, \textcite{Kunsch_1986} and \textcite{giraitis2001testing}.

A challenge in \textcite{iacone2022semiparametric} and \textcite{cuestas2016testing} lies in identifying the number of structural breaks and the order of the Chebyshev polynomial in the face of fractional integration, respectively. We present a solution to this challenge.  In particular, we combine the methods proposed by \textcite{cuestas2016testing} and \textcite{iacone2022semiparametric} by using a Chebyshev polynomial for modelling structural change and a semi-parametric setup for describing the short-run dynamics. We use a local Whittle objective for constructing a Lagrange multiplier (LM) test for fractional integration and for setting up a novel information criterion for selecting the order of the Chebyshev polynomial. 

We establish the asymptotic theory for the proposed testing procedure. In particular, we show that choosing too low an order for the Chebyshev polynomial leaves part of the smooth trend in the residuals, contaminates the spectrum at the origin and causes the LM-statistic to diverge under the null, rendering the resulting test uninformative. This low-frequency contamination is closely related to that discussed by \textcite{iacone2022semiparametric} for structural breaks. On the other hand, when the polynomial order is correctly specified or over-specified, the LM statistic based on the residuals have the same limiting behaviour as the infeasible procedures based on the true unobserved errors. More specifically, under alternatives drifting to the null at the standard local-to-null rate, the LM-statistic converges to a non-central chi-squared distribution, just as in the infeasible benchmark, cf.\ also \textcite{lobato1998nonparametric}, \textcite{shao2007local}, and \textcite{iacone2022semiparametric}. This implies that estimating the deterministic component does not entail any asymptotic loss of local power, provided that the selected polynomial order is not smaller than the true one. This asymptotic robustness to over-specification should not, however, be interpreted as saying that over-fitting is innocuous in practice. As noted by \textcite{bierens1997testing}, introducing superfluous Chebyshev terms may adversely affect finite-sample test performance, so that accurate order selection remains essential.

Indeed, our simulation evidence and empirical findings both indicate that the finite-sample consequences of over-specification can be substantial. We therefore theoretically analyse the behaviour of information criteria for selecting the order of the Chebyshev polynomial. We show that such conventional time-domain criteria as the BIC and HQ, which are valid in analogous settings with $I(0)$ errors, see \textcite{hall2013inference}, are generally unsuitable in the present environment. For, in the presence of positive long memory, they tend to interpret persistence as additional deterministic terms and therefore over-select the polynomial order. \textcite{lavielle2000least} consider penalty modifications to information criteria to account for persistence in the stochastic component. These modifications, however, are infeasible because the penalty depends on the unknown memory parameter. In addition, our simulation results indicate that these modified criteria may exhibit unsatisfactory finite-sample performance, underestimating for strong negative persistence and overestimating when persistence is strong. As a solution to these problems, we show that our novel frequency-domain local Whittle information criterion consistently selects the true polynomial order and thus provides a valid foundation for feasible inference on the fractional integration parameter.

We illustrate the methodology using the UK Great Ratios. \textcite{kaldor1961capital} first introduced the concept of Great Ratios, emphasising stable relationships among key macroeconomic variables. These ratios have since attracted substantial empirical interest, including recent analyses by \textcite{kapetanios2020time} and \textcite{chudik2023revisiting}, particularly regarding their integration properties. While \textcite{kapetanios2020time} find that UK Great Ratios display $I(1)$ behavior but can be represented as $I(0)$ processes when incorporating smooth deterministic trends, we extend this analysis to fractional integration. Our empirical findings reveal that the Great Ratios exhibit  long memory. Furthermore, we show that inference can be very sensitive to the choice of the polynomial order, underscoring the important role of the proposed local Whittle information criterion.

The structure of the remaining paper is as follows. Section \ref{sec:model} and \ref{sec:test-order-fract} introduce the model and the test of fractional integration we propose. Section \ref{information} studies the asymptotic properties of time-domain information criteria and introduces the new frequency-domain criterion for selecting the order of the Chebyshev polynomial. In Section \ref{s3} we conduct a simulation study to examine the finite sample properties of our proposed tests and our local Whittle information criterion. Section \ref{s4} presents an illustrative empirical example applying the methods developed in this article to analyse the UK Great Ratios. Section \ref{s5} concludes. Proofs of the theorems are relegated to the appendix while the supplement contains the results of an extensive Monte Carlo simulation study.

\section{Model specification, testing and order selection}\label{s2}

\subsection{The model}
\label{sec:model}

For the observable time series $y_t$, $t = 1,\ldots,T$, we consider a stationary fractional model supplemented by a Chebyshev polynomial, given by
\begin{align}
    y_t &= \sum_{n = 0}^k \beta_n P_{t}(n) + u_t, \label{eq1} \\
    u_t &= \Delta^{-\delta} \eta_t, \label{eq21}
\end{align}
with 
\begin{align}
    P_{t}(n) &= \sqrt{2} \cos \left(n \pi \frac{t - 0.5}{T} \right). \label{eq3}  
\end{align}
In this notation, each of the $k+1$ components $P_t(n)$ is referred to as \emph{Chebyshev function}, while the sum $\sum_{n=0}^k \beta_n P_t(n)$ will be called a \emph{Chebyshev polynomial} of order $k$. Note that
$P_t(0)$ is a constant term. The polynomial serves as a flexible smooth deterministic trend component that can approximate any square integrable and differentiable function of time arbitrarily closely, see \textcite{bierens1997testing}
and \textcite{bierens2010time}.

The so-called fractional integration operator in \eqref{eq21} is defined by $\Delta^{-\delta} = (1-L)^{-\delta} = \sum_{i=0}^{\infty} \pi_i(\delta) L^k $ with the binomial coefficients $\pi_0(\delta) = 1$ and
$\pi_i(\delta) = \delta (\delta + 1) \ldots (\delta + i -1)/ i !$ for $i \geq 1$, see for instance \textcite{hassler2019time}.  The error term $\eta_t$ is assumed to be a mean-zero stationary, causal non-linear process. Following \textcite{shao2007local,shao2007local1} and \textcite{iacone2022semiparametric} we impose the following assumptions on $\eta_t$, so as the be able to use the lemmas they establish for this class of processes. Let $\delta_0$ denote the true value of $\delta$.

\begin{assumption}\label{ass2}
    Let $$\eta_t = F(\ldots, \varepsilon_{t-1}, \varepsilon_t),$$ 
    where $\{\varepsilon_t\}_{t \in \mathbb{Z}}$ are independent and identically distributed (IID) random variables and $F$ is a measurable function such that $\eta_t$ is well-defined as a stationary, causal, ergodic process. For a random variable $\xi$ and $p > 0$, write $\xi \in L^p$ if $\|\xi\|_p = \left(E(\left|\xi\right|^p)\right)^{1/p} < \infty$. Let $\{\varepsilon^*_t\}_{t \in \mathbb{Z}}$ be an IID copy of $\{\varepsilon_t\}_{t \in \mathbb{Z}}$, $\mathscr{F}_t = (\ldots, \varepsilon_{t-1}, \varepsilon_t)$, $\mathscr{F}_0^* = (\mathscr{F}_{-1}, \varepsilon_0^*)$, $\eta_{\kappa}^* = F(\mathscr{F}_0^*, \varepsilon_1, \ldots, \varepsilon_{\kappa})$, and define the dependence measure $\theta_q({\kappa}) = \|\eta_{\kappa} - \eta_{\kappa}^*\|_q$, for some $q>\max\left\{4,\frac{1}{1+2\delta_0}\right\}$. Define the projection operator $\mathcal{P}_{\kappa} \xi = \mathbb{E}(\xi \mid \mathscr{F}_{\kappa}) - \mathbb{E}(\xi \mid \mathscr{F}_{{\kappa}-1})$. Then:\vspace{-3ex}
    \begin{enumerate}[noitemsep]
        \item[(i)] $\eta_t \in L^q$, $\sum_{\kappa_1,\kappa_2,\kappa_3}\left|\operatorname{cum}\!\left(\eta_0,\eta_{\kappa_1},\eta_{\kappa_2},\eta_{\kappa_3}\right)\right|<\infty$ and $\sum_{\kappa=0}^{\infty}\|\mathcal{P}_0\eta_{\kappa}\|_q<\infty$, where $\operatorname{cum}(\cdot)$ denotes the joint cumulant of its arguments. 
        \item[(ii)] $\sum_{{\kappa}=1}^{\infty} {\kappa} \theta_q({\kappa}) < \infty$. \label{a122}
        \item[(iii)] The spectral density of $\eta_t$, $f_{\eta}(\lambda)$, satisfies $f_{\eta}(\lambda) = G(1 + O(\lambda^2))$ as $\lambda \to 0^+$ for some $G \in (0,\infty)$.
    \end{enumerate}
\end{assumption}

Part $(i)$ imposes finite moments together with weak higher-order dependence through the cumulant and projection summability conditions. \textcite[Remark 3]{shao2007local} note that fourth-order cumulant summability is a standard condition in spectral analysis, while \textcite[Remark 2.5]{shao2007local1} show that, in the case of $F$ being a linear function, the corresponding projection and dependence conditions reduce to weighted summability restrictions on the linear coefficients. Part $(ii)$ captures how quickly the effect of a single shock dies out over time in terms of the dependence measure. As emphasised by \textcite[Remark 2.3]{shao2007local1}, this measure is especially appealing because it is directly tied to the data-generating mechanism. Part $(iii)$ requires the spectral density of $\eta_t$ to be smooth and strictly positive at the origin, so that $\eta_t$ behaves as a short-memory component in the local Whittle approximation, see \textcite[Remark 2.6]{shao2007local1}. Assumption \ref{ass2} covers a broad class of nonlinear models of $\eta_t$, including bilinear, threshold, GARCH, and ARMA-GARCH processes. The requirement on $q$ is needed to apply a functional central limit theorem later, see \textcite[Lemma 1]{iacone2022semiparametric}.

The parameter space for our model is defined as follows: $k \in \{0, 1, \ldots, K\}$, $\beta = (\beta_0, \beta_1, \ldots, \beta_k)' \in \mathbb{R}^{k+1}$ and $\delta \in (-0.5, 0.5)$. Finally, the true but unknown parameter values are denoted by $k_0$, $\beta_0 =  (\beta_{0,0}, \beta_{1,0}, \ldots, \beta_{k_0,0})' $, $\delta_0$, respectively.
The value $K$ represents the maximum order of the Chebyshev polynomial and must be larger or equal to the true value $k_0$. It can, in principle, be set arbitrarily large as long as it is less than $T-1$. In line with the literature, we impose the restriction $|\delta_0|<1/2$ , see e.g.\ \textcite{iacone2022semiparametric}. If $\delta_0$ were to exceed 0.5, the process $ u_t $ would become non-stationary and would dominate the deterministic component, entailing the
inconsistent estimation of the coefficients $\beta_n$, $n = 0, 1,\ldots,k$, of the Chebyshev functions, see e.g.\ \textcite{hualde2020truncated}. However, non-stationary data could be made stationary by first-differencing,
allowing us to still continue working within the framework of the model above.

The next two subsections contain the two main contributions of the paper. In Section \ref{sec:test-order-fract}, we develop an LM test for $H_0:\delta=\delta_0$ within the local Whittle framework and derive its asymptotic behaviour. This approach does not require preliminary estimation of $\delta$, nor does it require explicit modelling of the short-run dynamics, making it robust to misspecification of the latter,  see \textcite{shao2007local,iacone2022semiparametric,lobato1998nonparametric}. In Section \ref{information}, we then address the selection of the Chebyshev order $k$, since misspecification of the deterministic component adversely affects the testing problem. Because the LM test is itself based on the local Whittle objective, it is natural to base order selection on the same frequency-domain approach. This leads us to propose a novel local Whittle information criterion that provides a consistent basis for selecting $k$. We show that existing time-domain procedures, as opposed to that, are unsuitable in our setting.

\subsection{Testing the order of fractional integration}
\label{sec:test-order-fract}

We are interested in testing the null hypothesis
\begin{align}
    H_0 : \delta = \delta_0 \label{testing}
\end{align}
for $\delta_0 \in (-0.5, 0.5) $. \textcite{lobato1998nonparametric} study this problem in the context of model \eqref{eq1} when the Chebyshev polynomial only consists of a constant term,
i.e.\ when $k = 0$. They base the test on the Lagrange multiplier (LM) principle and the local Whittle objective function of \textcite{robinson1995gaussian}. The latter is given by
\begin{align}
    R_{u}(\delta;m) = \ln \left( \frac{1}{m} \sum_{j = 1}^m \lambda_j^{2\delta} I_u(\lambda_j) \right) - 2\delta \frac{1}{m} \sum_{j = 1}^m \ln(\lambda_j). \label{objectf}
\end{align}
Applying the LM principle to \eqref{objectf} yields the test statistic 
\begin{align*}
    \LM_u(\delta_0;m)
    =
    m \left[
    \frac{\partial^2 R_u(\delta;m)}{\partial \delta^2}
    \right]^{-1}
    \left[
    \frac{\partial R_u(\delta;m)}{\partial \delta}
    \right]^2
    \Bigg|_{\delta=\delta_0}. 
\end{align*}
Equivalently, the LM statistic can be expressed as the square of a $t$-type statistic, that is of interest in itself:
\begin{align}
    t_u(\delta_0;m) &= -\left( \frac{m^{-1/2} \sum_{j = 1}^{m} v_j \lambda_j^{2 \delta_0} I_{u} (\lambda_j) }{m^{-1} \sum_{j = 1}^{m}  \lambda_j^{2 \delta_0} I_{u}(\lambda_j)} \right), \label{test1}\\
    \LM_u(\delta_0;m) &= t_u(\delta_0;m)^2 \label{test2},
\end{align}
where $I_u(\lambda) = \left|w_u(\lambda)\right|^2$ is the periodogram of the error term $u_t$, $w_u(\lambda) = \frac{1}{\sqrt{2\pi T}} \sum_{t = 1}^T u_t e^{i \lambda t}$ is the discrete Fourier transform of $u_t$,
$\lambda_j = \frac{2\pi j}{T}$ are the Fourier frequencies, and $v_j = \ln(j) - \frac{1}{m} \sum_{j = 1}^m \ln(j)$, where $m = m(T)$ is the bandwidth. While the construction of the test statistic follows from \textcite{lobato1998nonparametric}, the large-sample conditions we impose are tailored to the broader nonlinear framework in Assumption \ref{ass2}. Accordingly, following \textcite{shao2007local} and \textcite{iacone2022semiparametric}, we impose the following conditions on the relative divergence rates of $m$ and $T$:
\begin{assumption}\label{bandw}
As $T \rightarrow \infty$, the bandwidth $m = m(T) \rightarrow \infty$ such that $T^{\epsilon}/ m + m/T^{2/3} \rightarrow 0$ for some fixed $\epsilon > 0$.
\end{assumption}
\textcite{lobato1998nonparametric} show that, when $\delta_0 = 0$ and under conditional homoskedasticity, the test statistics in \eqref{test1} and \eqref{test2} are asymptotically $\mathcal N(0,1)$ and $\chi^2_1$ under $H_0$, respectively. \textcite{shao2007local} extend these results to conditional heteroskedasticity. \textcite{iacone2022semiparametric} further extend the theory to any $|\delta_0|<1/2$.

As a first contribution of this paper, we consider the model in \eqref{eq1}-\eqref{eq21}  when the Chebyshev polynomial comprises not only a constant term. We solve the problem of testing $H_0$ in \eqref{testing} when $u_t$ is unobserved, as will in practice usually be the case.  In essence, we replace the unobservable $u_t$ in
\eqref{eq1} by its sample counterpart, estimating $\beta$ by ordinary least squares (OLS) for a given value of $k$. Neglecting the short-run dynamics in $u_t$ is of course inefficient yet, under the assumptions specified above, the
OLS estimator is consistent. The fact that the true order $k_0$ of the Chebyshev polynomial is in fact unknown makes the tests depend on whether $k$ is under- or overspecified in the model. This problem will be remedied in Section
\ref{information} below, where an information criterion is suggested for the consistent selection of $k$.

For a given $k$, let $y = (y_1,\ldots,y_T)'$, $x_t(k)$ $=$ $(P_t(0),\ldots, P_t(k))'$ and $X(k) = (x_1(k),\ldots,x_T(k))'$. Then the OLS estimator of $\beta$ is given by
$\hat{\beta}(k)= (X(k)'X(k))^{-1}X(k)'y$. Define the corresponding residuals,
\begin{align}
    \hat{u}_t(k) = y_t - x_t(k)' \hat{\beta}(k). \label{res}
\end{align} 
The statistics in \eqref{test1} and \eqref{test2} based on $\hat{u}_t(k)$ instead of $u_t$ are denoted by $ t_{\hat{u}(k)}(\delta;m)$ and $\LM_{\hat{u}(k)}(\delta;m)$, respectively. In the remainder of this section, we will
establish the limiting behaviour of $t_{\hat{u}(k)}(\delta;m)$ and $\LM_{\hat{u}(k)}(\delta;m)$, with Theorem \ref{thm1} covering the under-specified case ($k < k_0$) and Theorem \ref{thm2} addressing the correctly specified
($k = k_0$) and over-specified ($k > k_0$) cases. These asymptotic results hinge on the fact that the coefficients $\beta_n$, $n = 0, 1,\ldots,k$, of the Chebyshev functions can be consistently estimated at rate
$T^{1/2-\delta_0}$, even when $k$ is under- or over-specified, cf.\ Lemma \ref{consCheby}. This is achieved via the orthogonal property of the Chebyshev polynomial and the application of the functional central limit theorem, which, as mentioned earlier, holds in our setting.

The following theorem shows the limiting behaviour of $t_{\hat{u}(k)}(\delta;m)$ and  $\LM_{\hat{u}(k)}(\delta;m)$ under $H_0$ when $k$ is under-specified, under an additional restriction on the bandwidth $m$. 
\begin{theorem} \label{thm1}
Let $y_t$ be generated according to \eqref{eq1}-\eqref{eq21} and let Assumption \ref{ass2} hold.
The bandwidth $m$ satisfies the conditions in Assumption \ref{bandw} and $T^{1-2\delta_0}   \ln(m) m^{- 1/2} \rightarrow \infty$. Then for any $k < k_0 $ with $k_0 > 0$ and any $\delta_0 \in (-0.5,0.5)$,
\begin{align*}
    t_{\hat{u}(k)}(\delta;m) &\overset{p}{\rightarrow} \infty, \\
   \LM_{\hat{u}(k)}(\delta;m) &\overset{p}{\rightarrow} \infty,
\end{align*}
under $H_0 : \delta =  \delta_0$. If, moreover, $T^{1-2\delta_0} m^{-1} \rightarrow \infty$, then
\begin{align*}
    t_{\hat{u}(k)}(\delta;m) &\overset{p}{\sim}  C \sqrt{m} \ln(m), \\
   \LM_{\hat{u}(k)}(\delta;m) &\overset{p}{\sim} C^2 m\ln^2(m),
\end{align*}
under $H_0 : \delta =  \delta_0$, where $C$ is a positive finite constant.
\end{theorem}

Three observations can be made about Theorem \ref{thm1}: First, if the DGP includes Chebyshev functions beyond the constant term, i.e.\ if $k_0 > 0$,  yet these functions are ignored in the
model by setting $k = 0$ then both test statistics will diverge under the null. Therefore, in this case, both tests are not informative. Secondly, even if Chebyshev functions beyond the intercept are included in the model ($k > 0$) but the order of the Chebyshev
polynomial in the model is smaller than the true number ($k < k_0$), the test statistics will still diverge. Thirdly, the rate of divergence of both statistics in the second part of Theorem \ref{thm1} is the same
as in the setting studied by \textcite{iacone2022semiparametric}, who address level breaks instead of smooth trends. Hence, neglecting smooth trends and neglecting level breaks affects these tests through a similar low frequency contamination mechanism.

The assumptions on the bandwidth in Theorem \ref{thm1} are best interpreted as relative strength conditions: they guarantee divergence of the test statistics by making the contribution of the omitted Chebyshev component dominate the error term, cf.\ \textcite{mccloskey2013memory}. The weaker restriction in the first part of the theorem is sufficient to make the contamination component dominate in the numerator of the $t$-type statistic, which implies divergence of both test statistics. The stronger restriction in the second part is imposed to ensure that the omitted component also dominates the denominator, which is what delivers the explicit divergence rates in the theorem.

Essentially, therefore, Theorem \ref{thm1} says that both the $t$- and the LM-test are uninformative under the null when the Chebyshev polynomial is underspecified. Theorem \ref{thm2} rectifies this by establishing that both tests are well-behaved under the null and under local alternatives as along as  the Chebyshev polynomial is correctly or overspecified relative to the true value $k_0$. This result highlights the robustness of the proposed testing procedures to potential overfitting in the model and ensures that the limiting distributions remain valid provided that $k \geq k_0$.

\begin{theorem} \label{thm2}
Let $y_t$ be generated according to \eqref{eq1}-\eqref{eq21} and let Assumption \ref{ass2} hold.
The bandwidth $m$ satisfies the conditions in Assumption \ref{bandw}. Then for any $k \geq k_0$ and any $\delta_0 \in (-0.5,0.5)$,
\begin{align}
      t_{\hat{u}(k)}(\delta_0;m) &\overset{d}{\rightarrow} \mathcal N(2c,1), \\
   \LM_{\hat{u}(k)}(\delta_0;m) &\overset{d}{\rightarrow} \chi_1^2(4c^2),
\end{align}
under $H_c :\delta = \delta_0 + c m^{-1/2}$.
\end{theorem}
The theorem establishes that, under local alternatives converging to the null at the $\sqrt{m}$-rate, the $t$-statistic converges in distribution to a normal distribution with mean $2c$ and variance one, while the LM-type
statistic converges to a non-central chi-square distribution with one degree of freedom and non-centrality parameter $4c^2$. These limiting distributions are identical to those obtained using the infeasible versions of the
statistics based on the true errors $u_t$, as shown in \textcite{iacone2022semiparametric}. This equivalence implies that, asymptotically, the estimation of $u_t$ by $\hat{u}_t(k)$ does not lead to any loss of local power, provided that
the Chebyshev order is sufficiently large. Thus, the proposed procedures are not only consistent but also asymptotically efficient in the sense that they achieve the same local asymptotic power as the infeasible tests based on the
true errors $u_t$. Note that Assumption \ref{bandw} on the bandwidth is sufficient for the asymptotic theory in Theorem \ref{thm2}.

\subsection{Selecting the Chebyshev order}
\label{information}

A crucial issue in the testing procedures suggested in Section \ref{sec:test-order-fract} is of course the fact that the true $k_0$ is unknown. From Theorem \ref{thm1}, it follows that, if $k < k_0$, both statistics diverge under
the null hypothesis. On the other hand, Theorem \ref{thm2} shows that if $k \geq k_0$, the tests have a well-defined asymptotic distribution under both the null and local alternatives. Therefore, the choice of $k$ is
important. \textcite{bierens1997testing}, too, emphasises the importance of $k$, noting that the power and size of the unit root test he examines is highly sensitive to the choice of $k$. The literature contains a number of procedures for selecting $k$, three of which will be discussed in the next
subsection. We will derive their properties and argue that they all have drawbacks. We subsequently suggest a new solution to the selection problem.

\subsubsection*{Existing selection procedures}

First, \textcite{cuestas2016testing} propose a method to address this challenge by selecting $k$ via a sequential testing procedure. Their general-to-specific approach starts with a high-order polynomial and gradually removes statistically insignificant
Chebyshev coefficients based on their $t$-statistics, continuing until only significant coefficients remain. However, the method has limitations: First, the distribution of the estimated Chebyshev coefficients depends on the unknown
true fractional integration order $\delta_0$ which must be estimated to perform the $t$-tests, cf.\ Lemma \ref{consCheby}. Additionally, the long-term variance estimator required to studentise the coefficient estimates typically also depends on $\delta_0$. A detailed discussion is provided in \textcite{abadir2009two}.

Information criteria offer an alternative approach for conducting inference about the order of the Chebyshev polynomial. These criteria are frequently employed in the context of structural breaks in models with $I(0)$ errors, see
\textcite{bai2003computation} and \textcite{hall2013inference}, as well as in models with $I(\delta)$ errors where $|\delta| < 1/2$, see \textcite{lavielle2000least}. The criteria in this literature all take the form 
\begin{align}
    \IC(k) &= B(k) + (k+1)A(T)  \label{ICapproach}   
\end{align}
where $B(k)$ is some estimate of the in-sample fit and $A(T)$ is a positive deterministic penalty term. The estimate of the order of the Chebyshev polynomial, denoted $\hat{k}$, is the value
that minimises $\IC(k)$, that is,
\begin{align}
    \hat{k} = \argmin_{k \in \{0,1,\ldots,K\}} \IC(k). \label{k} 
\end{align}

As the second of the selection procedures we discuss, consider the most natural specifications of $B(k)$ and $A(T)$, viz.\ the Bayesian Information Criterion and the Hannan-Quinn information criterion. We refer to these conventional time-domain criteria as BIC and HQ. These criteria both set $B(k) = \ln \hat{\sigma}^2(k)$ where 
\begin{align}
    \ln \hat{\sigma}^2_u (k) = \ln\ \left( \frac{1}{T} \sum_{t = 1}^T \hat{u}^2_t(k) \right) \label{eq:sighatu}
\end{align}
with $\hat{u}_t(k)$ defined in \eqref{res}, while their penalty terms are respectively, $A(T) = \log(T)/T$ and $A(T) = 2c\log(\log(T))/T$, where $c > 1$. It can be verified that these penalty terms satisfy the following assumption:
\begin{assumption} \label{ass4}
As $T \rightarrow \infty$, $A(T) = o(1)$ and $TA(T)\rightarrow \infty$. If $\delta_0 > 0 $ assume that, in addition, $T^{1-2 \delta_0} A(T)= o(1)$.
\end{assumption}
This assumption is akin to that in \textcite{hall2013inference} and stipulates that the rate at which $A(T)$ converges to zero must be at most $T$ and, if $\delta_0 > 0$, at least $T^{1 - 2\delta_0}$. While, in linear models
with structural breaks and $I(0)$ errors, \textcite{hall2013inference} show that a penalty term $A(T)$ satisfying Assumption \ref{ass4} yields a consistent estimate of the number of breaks, we show in the following theorem that,
in our model with $I(\delta)$ errors, the consistency of $\hat{k}$ depends on $\delta_0$.
\begin{theorem}  \label{l1}
Let $y_t$ be generated according to \eqref{eq1}--\eqref{eq21} and let Assumptions \ref{ass2} and \ref{ass4} hold.
If $\delta_0 > 0$, then $P(\hat{k} > k_0) \rightarrow 1$. If $\delta_0 \leq 0$, then $\hat{k} \overset{p}{\rightarrow} k_0$.
\end{theorem}
This theorem shows that $\hat{k}$ leads to overestimation when $\delta_0 > 0$, implying that long memory is misinterpreted as Chebyshev functions. As noted by \textcite{bierens1997testing}, such overestimation is undesirable
because it introduces unnecessary Chebyshev functions into the model. This, in turn, distorts significantly the size and power of the tests in Section \ref{sec:test-order-fract}, as the inclusion of superfluous Chebyshev functions 
adds noise and does not improve the model's accuracy. This issue is further demonstrated in the simulation study in Section \ref{s3}. However, Theorem \ref{l1} also shows that $\hat{k}$ is consistent when $\delta_0 \leq 0$,
confirming that the analogous result of \textcite{hall2013inference} for models with structural breaks and $I(0)$ errors also holds for some parameter values in our setting.

The reason for the overestimation of the Chebyshev polynomial order when $\delta_0 > 0$ is that the BIC, the HQ and other criteria satisfying Assumption \ref{ass4} do not impose a sufficiently strong penalty when persistence is
present in the time series. In the context of structural breaks, this has motivated \textcite{lavielle2000least} to modify the penalty term $A(T)$ such that it converges to zero at a slower rate than in Assumption \ref{ass4} when
$\delta_0 > 0$. Following \textcite[Theorem 9]{lavielle2000least}, we thus replace Assumption \ref{ass4} by 
\begin{assumption} \label{ass3}
As $T \rightarrow \infty$, $A(T) = o(1)$ and $T^{1-2\delta_0}A(T) \rightarrow \infty$, for any $\delta_0 \in (-0.5, 0.5)$. 
\end{assumption}
Consequently, $A(T)$ must now converge to zero at a slower rate than $T^{1 - 2\delta_0}$. In particular, \textcite{lavielle2000least} modify the BIC penalty term by making the rate in the denominator depend on $\delta_0$,
defining $A(T) = \log(T)/T^{1-2\delta_0}$. Similarly, one could modify the HQ criterion by setting $A(T) = 2c\log(\log(T))/T^{1-2\delta_0}$. We refer to these infeasible $\delta_0$-dependent information criteria, the third procedure type in our categorisation, as $\delta$-BIC and $\delta$-HQ.
With $B(k)$ still given by $\ln \hat \sigma^2_u (k)$ in \eqref{eq:sighatu}, we
can then prove the following behaviour of $\hat k$ in \eqref{k}:
\begin{theorem}  \label{thm3}
Let $y_t$ be generated according to \eqref{eq1}--\eqref{eq21} and let Assumptions \ref{ass2} and \ref{ass3} hold. Then, $\hat{k} \overset{p}{\rightarrow} k_0$. 
\end{theorem}
 Theorem \ref{thm3} shows that the estimated order of the polynomial is consistent when $A(T)$ satisfies Assumption \ref{ass3}. The modification of the penalty of course poses a challenge: since $\delta_0$ is unknown the $\delta$-BIC and $\delta$-HQ cannot be computed in practice. 

All three procedures for determining $k$ described in this section are thus dependent on $\delta_0$, making them ineffective for the use with the tests proposed in Section \ref{sec:test-order-fract}. In the following, we will suggest a solution out of this impasse.

\subsubsection*{A new selection procedure}

As a second contribution of this paper, we now introduce a new selection procedure for $k$ to avoid the dependence on $\delta_0$ of the information criteria discussed above. Relative to those approaches, it re-defines both the
estimator $B(k)$ and the penalty term $A(T)$ in \eqref{ICapproach}, taking into account the estimation of the memory parameter $\delta$ as well as the non-parametric nature of the short-run dynamics. The main idea is to move order selection to the frequency domain and use the local Whittle (LW) objective function as the measure of fit.

Specifically, an estimator of $B(k)$ for a given candidate model with $k$ Chebyshev functions is obtained by a 2-step procedure. In the first step, the long-memory parameter is estimated semi-parametrically using the local Whittle estimator $\hat \delta (k)$ applied to the residuals $\hat{u}_t(k)$ defined in \eqref{res}, i.e.\ 
\begin{align*}
    \hat\delta(k)=\arg\min_{\delta \in\mathcal D} R_{\hat{u}(k)} (\delta;m)
\end{align*}
where $m = m(T)$ again satisfies the bandwidth rate condition in Assumption \ref{bandw}. The parameter space is $\mathcal D = [ \Delta_1,\Delta_2 ]$, with $-1/2 < \Delta_1 < \Delta_2 < 1/2$, $\delta_0 \in \mathcal D$. 
In the second step, define the goodness-of-fit term $B(k)$ in \eqref{ICapproach} as
\begin{align}
    B(k) = R_{\hat{u}(k)} (\hat\delta(k);m).\label{eq:sighateps}
\end{align}
i.e.\ the local Whittle objective in \eqref{objectf} evaluated at the first-step estimate. 

Turn now to the penalty term $A(T)$ in \eqref{ICapproach}. In contrast to Assumptions \ref{ass4} and Assumption \ref{ass3} above, it is defined in terms of the bandwidth $m$ used by the first step local Whittle estimator of $\delta$:
\begin{assumption} \label{ass6}
As $T \rightarrow \infty$, $A(T) = o(1)$ and $A(T) m (T)\rightarrow \infty$. 
\end{assumption}
According to Assumption \ref{ass6}, the penalty $A(T)$ should decay to zero sufficiently fast in order to avoid under-selecting the
Chebyshev order, while not decaying too quickly to prevent over-selection. Natural choices for $A(T)$ mirror the standard BIC or HQ penalties with the effective sample size $m = m(T)$ in place of $T$, namely
\begin{align}
A(T) &= \frac{\log\left( m (T) \right)}{ m (T)} \label{option1} \\
A(T) &= \frac{2c \log\left(\log\left(m (T) \right)\right)}{m (T)},\label{option2}
\end{align}
with $c > 1$. We refer to the resulting criteria as LW-BIC and LW-HQ, respectively.

Given a model with $k$ Chebyshev functions, we hence base the information criterion on the local Whittle fit term $R(\hat{\delta}(k);m)$ in \eqref{eq:sighateps} and on the penalty $A(T)$ in \eqref{option1} or
\eqref{option2}, such that the optimal polynomial order is selected as in \eqref{k}.

The following theorem shows the consistency of $ \hat{k}$.
\begin{theorem} \label{thm5} Let $y_t$ be generated according to \eqref{eq1}-\eqref{eq21} and let Assumptions \ref{ass2}, \ref{bandw} and \ref{ass6} hold. Furthermore, assume that $P(\hat{\delta}(k)\leq \delta_0) \rightarrow 0$  holds  for all $k < k_0$. Then
\begin{align*}
\hat{k} \overset{p}{\to} k_0.
\end{align*}
\end{theorem}

The mechanism behind  Theorem \ref{thm5} is straightforward. When $k<k_0$, the low-frequency contamination left in $\hat u_t(k)$ worsens the local Whittle fit. As a result, $B(k)>B(k_0)$ with probability tending to one. Since $A(T)=o(1)$, the penalty component $(k+1)A(T)$ is asymptotically negligible relative to this deterioration in fit, and the criterion rules out underfitting. When $k> k_0$, the contamination is removed and $\hat\delta(k)$ is consistent with the $\sqrt{m}$-usual rate, cf.\ Lemma \ref{thm6}. Yet, the resulting difference in fit components is asymptotically negligible, i.e.\ $B(k)-B(k_0)=O_p(m^{-1})$, and no longer separates correctly-specified from overspecified models. Assumption \ref{ass6} is then intrumental: $A(T)m\to\infty$ ensures that the penalty term dominates the $O_p(m^{-1})$ fit differences, thereby preventing over-selection. 

The additional condition $P(\hat{\delta}(k)\leq \delta_0) \rightarrow 0$ for all $k < k_0$ requires that, should the Chebyshev polynomial be underspecified, the estimator of $\delta$ is sufficiently large and thereby compensates for the low frequency contamination in $\hat{u}(k)$. If this spurious long memory is not present then the local Whittle goodness-of-fit term in \eqref{eq:sighateps} may fail to penalise under-specified models. In that case, the information criterion may favour $k < k_0$, leading to under-selection. Assumptions of this type are standard in related problems, see e.g.\  \textcite{qu2011test} and \textcite{SibbertsenEtAl18}.

While we use the local Whittle estimator in the first step for logical consistency, the consistency result in Theorem \ref{thm5} does in fact not hinge on this particular choice. More generally, our consistency proof uses only two properties of the first–step estimator $\hat\delta(k)$. First, under under-fitting ($k<k_0$), it yields spurious long memory in the sense that $P(\hat{\delta}(k)> \delta_0)\rightarrow 1$. Second, under correct specifcation or over-fitting ($k\geq k_0$), it is $\sqrt{m}$-consistent, i.e.\ $m^{1/2} (\hat{\delta}(k)-\delta_0) = O_p(1)$. Consequently, the local Whittle estimator can be replaced by any other semi-parametric estimator that satisfies these two conditions. Examples may include the exact local Whittle estimator of
\textcite{exactShimotsu} and the fully extended local Whittle estimator of \textcite{abadir2007nonstationarity}.  
Similarly, we use the same notation $m$ for the value of the bandwidth in the construction of both the test statistics in \eqref{test1}-\eqref{test2} and the LW information criteria. These quantities could in principle, however, be based on different bandwidths, provided that they each meet Assumption \ref{bandw}.

\section{Monte Carlo simulation}\label{s3}

In this section, we investigate the finite-sample properties of the tests discussed in Section \ref{s2}. 
Section \ref{s3a} considers a setting in which the order of the Chebyshev polynomial is fixed at some value $k$ that is not necessarily the true order. Subsequently, Section \ref{s3b} evaluates the tests when the order is selected via the information-criterion approaches described in Section \ref{information}, which is more realistic for empirical work. 
We conclude Section \ref{s3b} by a set of practical recommendations based on the simulation evidence reported in the Supplementary Appendix. 
All computations are performed in MATLAB 2019a, see \textcite{MATLAB}.

\subsection{Tests with a fixed Chebyshev order}\label{s3a}

\begin{figure}[!htbp]
    \centering
\subfloat[$T = 256$,  $m =  \left \lfloor T^{0.65} \right \rfloor$]{
  \includegraphics[width=0.50\textwidth]{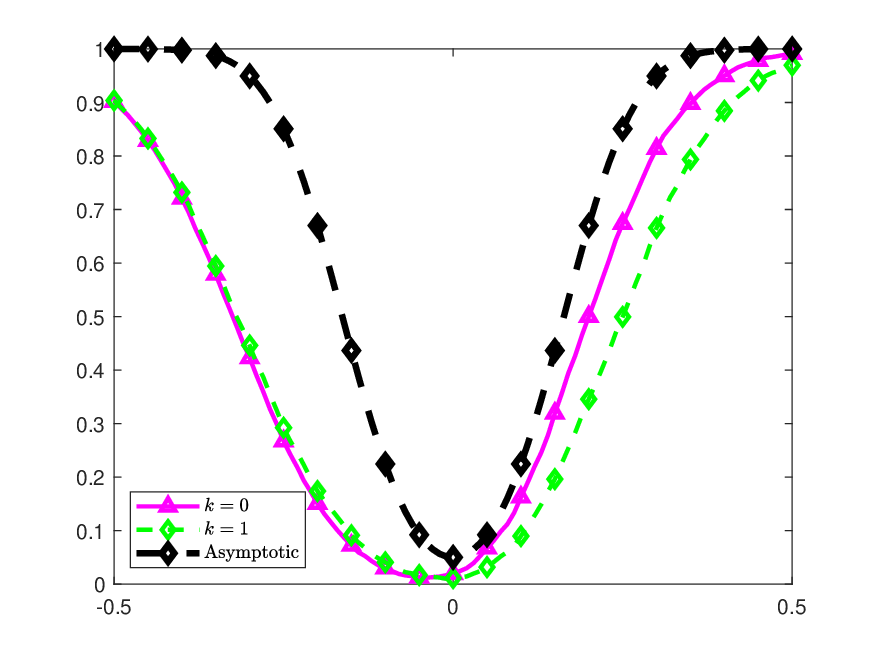}
}
\subfloat[$T = 512$,  $m =  \left \lfloor T^{0.65} \right \rfloor$]{
  \includegraphics[width=0.50\textwidth]{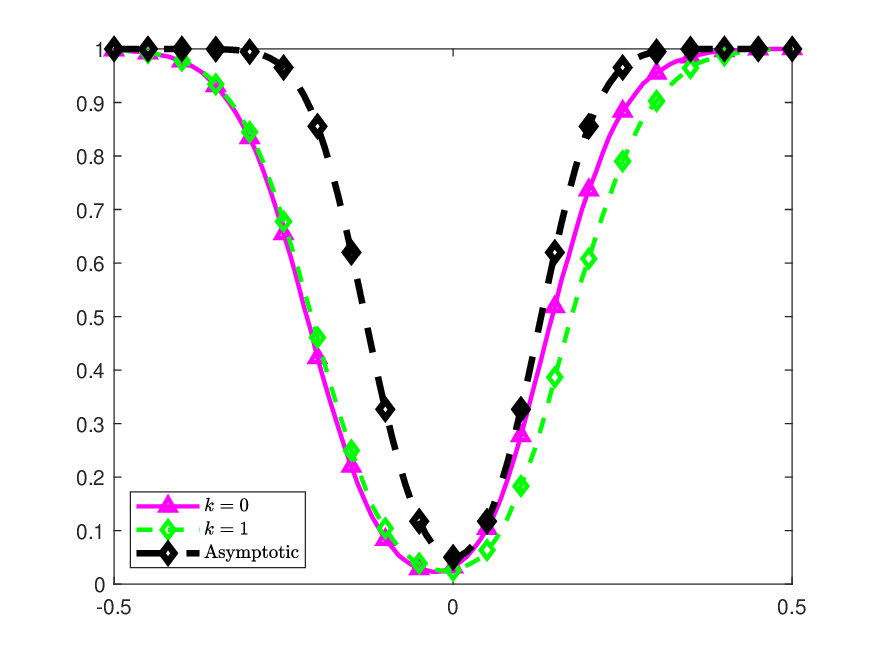} 
}
\hspace{0mm}
\subfloat[$T = 256$,  $m =  \left \lfloor T^{0.80} \right \rfloor$]{
  \includegraphics[width=0.50\textwidth]{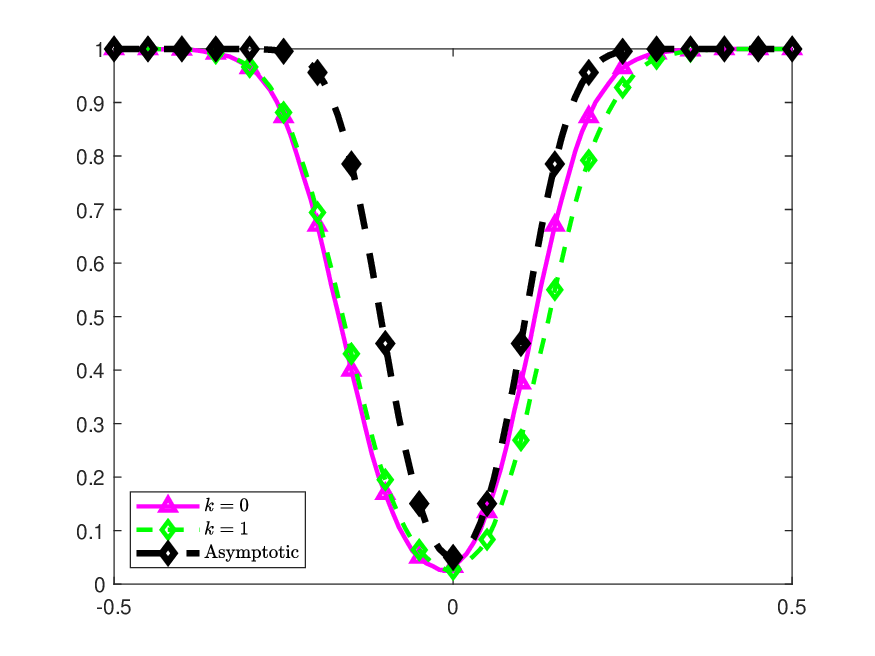}
}
\subfloat[$T = 512$,  $m =  \left \lfloor T^{0.80} \right \rfloor$]{
  \includegraphics[width=0.50\textwidth]{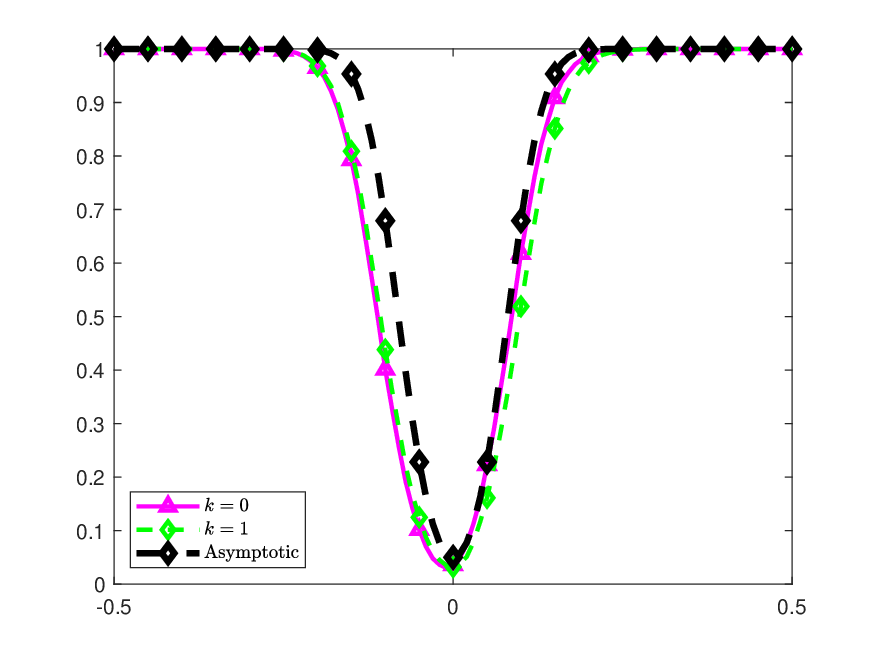}  
}
    \caption{Simulated rejection frequencies of $H_0 : \delta = 0$, averaged over 5,000 replications. The DGP sets $k_0 = 0$ with $\beta_{0,0} = 0$, while $\epsilon_t \sim \text{NIID}(0,1)$ and $\delta_0 \in [-0.499, 0.499]$. The model is given by $y_t = \beta_0 +  \Delta^{-\delta} \epsilon_t$ when $k = 0$, and by $y_t = \beta_0 + \beta_1 P_t(1) + \Delta^{-\delta} \epsilon_t$ when $k = 1$. The LM-statistic is based on the OLS residuals $\hat u_t (k)$ in \eqref{res}.}
    \label{f1}%
\end{figure}

Figure \ref{f1} compares the asymptotic behaviour of the LM-type statistic as predicted by Theorem \ref{thm2} with finite-sample simulations. The DGP  given in \eqref{eq1}--\eqref{eq21} is assumed to contain a constant term only, so the true order of the Chebyshev polynomial is $k_0 = 0$, and we set $\beta_{0,0} = 0$ without loss of generality. The short-run dynamics are generated by $\eta_t \sim \text{NIID}(0,1)$, while the fractional parameter is $\delta_0 \in [-0.499, 0.499]$. The baseline setup of the simulation is a sample size $T = 512$ and a bandwidth $m =  \left \lfloor T^{0.65} \right \rfloor$. Following the literature, we also consider a larger bandwidth of $m =  \left \lfloor T^{0.80} \right \rfloor$. So as to match the empirical sample in Section \ref{s4}, we examine a sample size of $T = 256$, too.  The null hypothesis of interest is $H_0 : \delta = 0$ and the nominal asymptotic significance level is 5\%. Rejection frequencies of $H_0$ are reported as a function of $\delta_0$, averaged over 5,000 replications. The LM-type statistic $\LM_{\hat u \left( k \right)} $ is based on the model whose Chebyshev polynomial is either correctly specified ($k = 0$) and overspecified ($k = 1$).  In addition, the asymptotic local power curve from Theorem \ref{thm2} is shown as a benchmark. 

For all values of $T$ and $m$, the rejection frequencies under $H_0$ approximate the nominal size well. Simulated power is almost indistinguishable between the correctly specified and the overspecified model for negative values of $\delta_0$. For positive values of $\delta_0$, power in the correctly specified model is uniformly higher than in the overspecified model. It is closer to the asymptotic benchmark for positive than for negative values of $\delta_0$. The difference between the correctly specified and overspecified models as well as the difference to the asymptotic benchmark shrink with larger samples and higher bandwidths. Overall, the finite sample patterns match the asymptotic predictions of Theorem \ref{thm2}, especially for larger bandwidths.

\begin{figure}[!htbp]
    \centering
   
\subfloat[$T = 256$,  $m =  \left \lfloor T^{0.65} \right \rfloor$]{
  \includegraphics[width=0.50\textwidth]{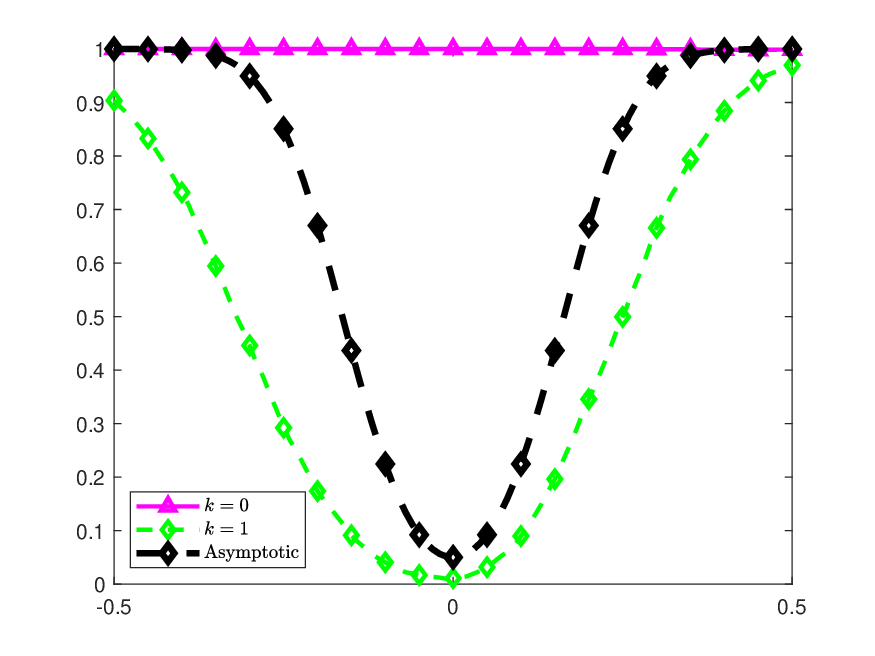}
}
\subfloat[$T = 512$,  $m =  \left \lfloor T^{0.65} \right \rfloor$]{
  \includegraphics[width=0.50\textwidth]{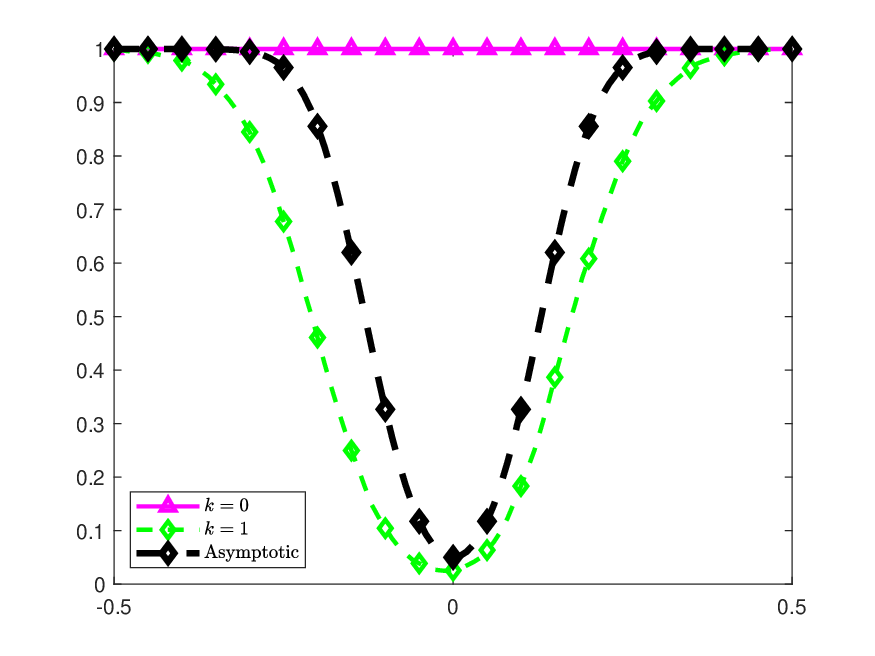} 
}
\hspace{0mm}
\subfloat[$T = 256$,  $m =  \left \lfloor T^{0.80} \right \rfloor$]{
  \includegraphics[width=0.50\textwidth]{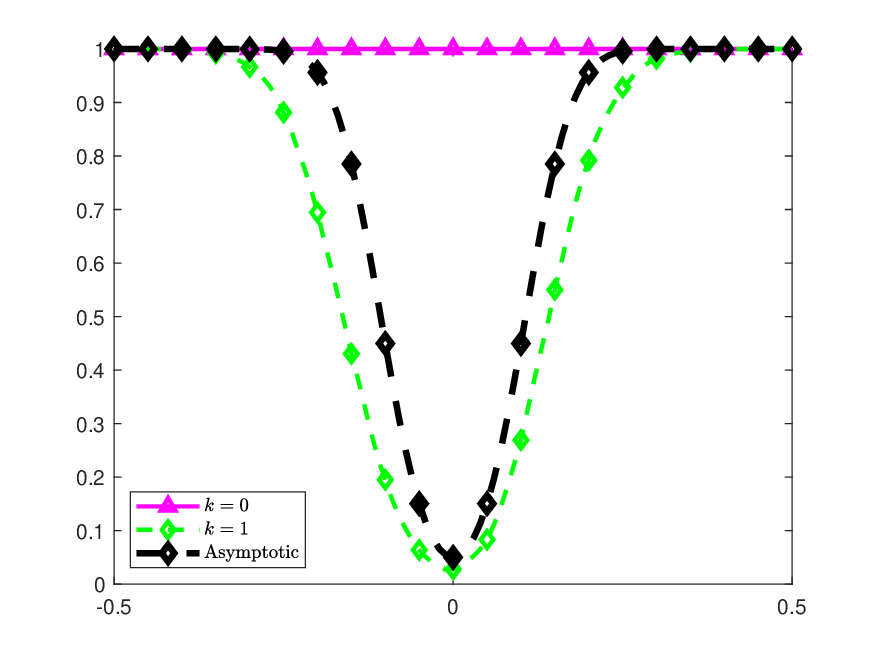}
}
\subfloat[$T = 512$,  $m =  \left \lfloor T^{0.80} \right \rfloor$]{
  \includegraphics[width=0.50\textwidth]{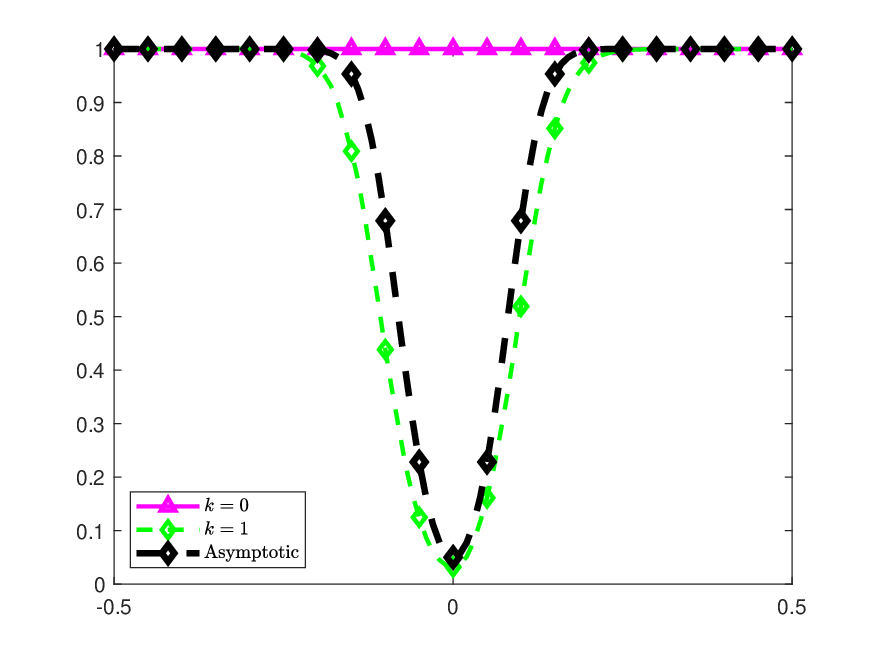}  
}
    \caption{ Simulated rejection frequencies of $H_0 : \delta = 0$, averaged over 5,000 replications. The DGP sets $k_0 = 1$ with $\beta_{0,0} = 0$ and $\beta_{1,0} = 1$, while $\epsilon_t \sim \text{NIID}(0,1)$ and $\delta_0 \in [-0.499, 0.499]$. The model is given by $y_t = \beta_0 +  \Delta^{-\delta} \epsilon_t$ when $k = 0$, and by $y_t = \beta_0 + \beta_1 P_t(1) + \Delta^{-\delta} \epsilon_t$ when $k = 1$. The LM-statistic is based on the OLS residuals $\hat u_t (k)$ in \eqref{res}.}%
    \label{f2}%
\end{figure}

Figure \ref{f2} extends the simulation evidence to include an illustration of Theorem \ref{thm1}. The DGP is as for Figure \ref{f1}, yet the true order of the Chebyshev polynomial is now $k_0 = 1$, so the DGP contains an intercept and one additional Chebyshev function. The true coefficient values are set to $\beta_{0,0} = 0$ and $\beta_{1,0} = 1$, respectively. The LM-type statistic $\LM_{\hat u \left( k \right)} $ is again based on the model whose Chebyshev polynomial is of order $k = 0$ or $k = 1$, meaning that the model is now underspecified or correctly specified, respectively. 

A clear distinction is now apparent between these two model specifications. When $k = 1$ such that the model is correctly specified, the finite-sample size and power of the LM-type test mirror the corresponding conclusions in Figure \ref{f1}: The simulated size is close to the nominal size for all values of $T$ and $m$, while the simulated power is substantial and approaches its asymptotic benchmark for increasing sample sizes and bandwidths.
However, when $k=0$ such that the model is underspecified the test rejects 100\% of the time 
when $\delta_0 = 0$, confirming the prediction of Theorem \ref{thm1}. Indeed, the test always rejects when the null is not true.

\subsection{Tests with Chebyshev order selected by information criteria}\label{s3b}

We now examine the behaviour of our LM test when the order of the Chebyshev polynomial is selected by the information criteria  discussed in Section \ref{information}. Specifically, we consider {\it (i)} the standard time-domain HQ criterion, {\it (ii)} the infeasible $\delta$-HQ criterion proposed by \textcite{lavielle2000least} and {\it (iii)} our local Whittle HQ information criterion LW-HQ. In all cases, we set the penalty tuning constant in \eqref{option2} equal to $c=1.0001$. The experimental setup follows that of the previous section, involving a Chebyshev polynomial of true order $k_0=1$ with coefficient $\beta_{0,0}=0$ and $\beta_{1,0}=1$. We omit the case $k_0=0$ since it leads to qualitatively identical conclusions. The true fractional parameter is $\delta_0 \in \mathcal D$, with $\mathcal D = [-0.499, 0.499]$. The LM-type statistic $\LM_{\hat u \left( k \right)} $ for a model with correctly specified polynomial order $k = k_0$ is included as a benchmark.  The parameter space to estimate $\delta$ over is taken to be $\mathcal D$. The LW-HQ criterion is computed using the bandwidth $m=\lfloor T^{0.65}\rfloor$. 

\begin{figure}[!htbp]
    \centering
   
\subfloat[$T = 256$,  $m =  \left \lfloor T^{0.65} \right \rfloor$]{
  \includegraphics[width=0.50\textwidth]{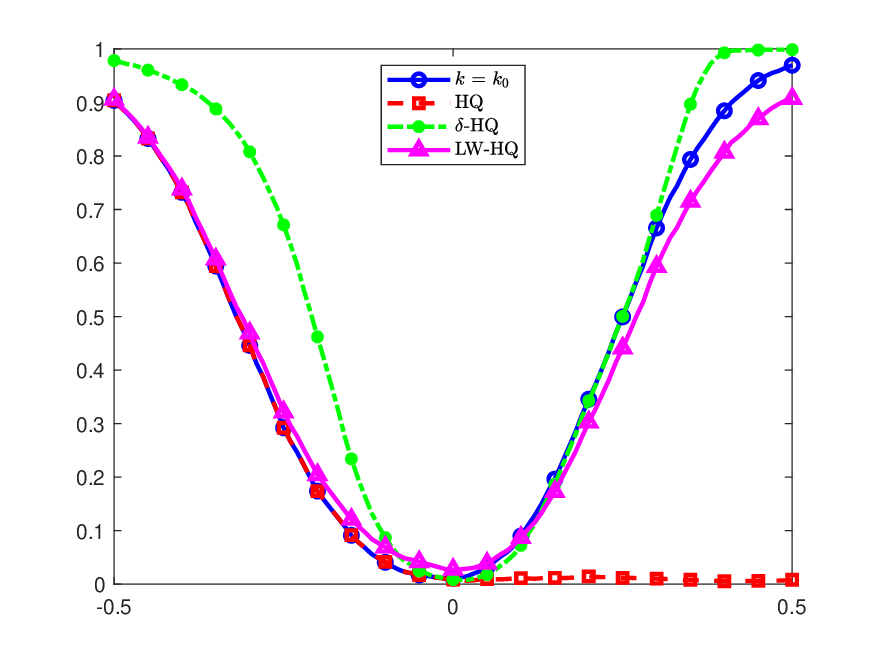}
}
\subfloat[$T = 512$,  $m =  \left \lfloor T^{0.65} \right \rfloor$]{
  \includegraphics[width=0.50\textwidth]{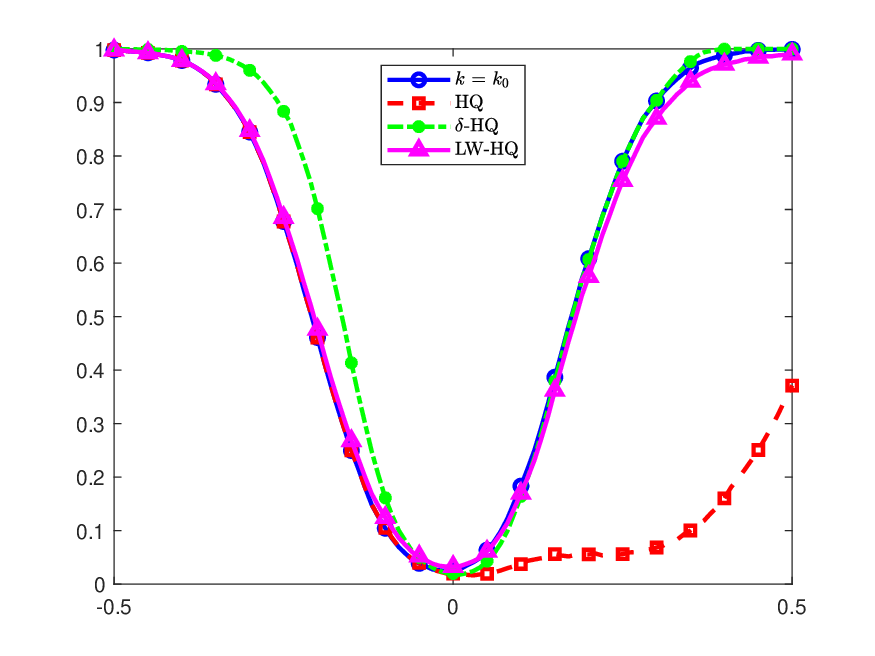} 
}
\hspace{0mm}
\subfloat[$T = 256$,  $m =  \left \lfloor T^{0.80} \right \rfloor$]{
  \includegraphics[width=0.50\textwidth]{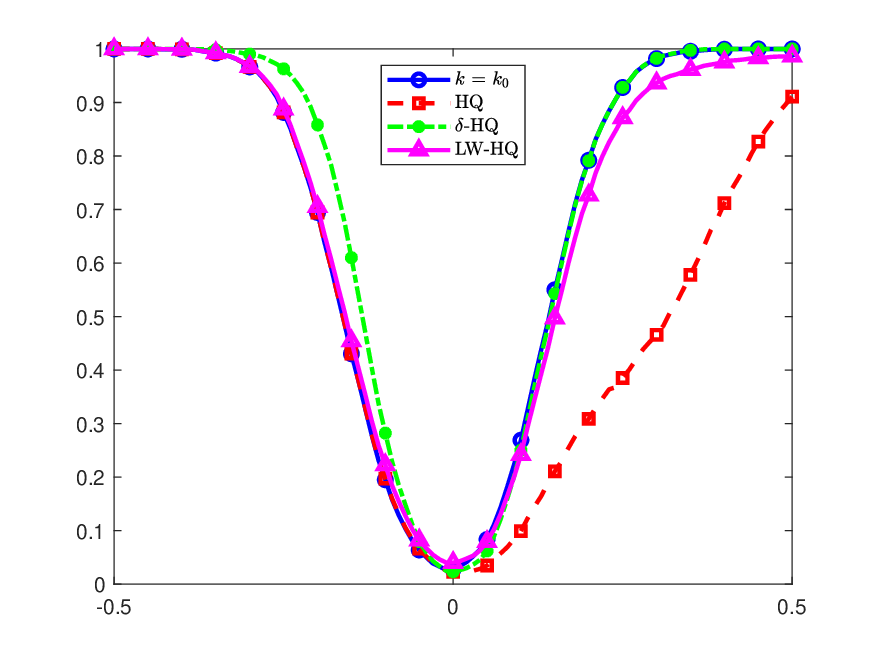}
}
\subfloat[$T = 512$,  $m =  \left \lfloor T^{0.80} \right \rfloor$]{
  \includegraphics[width=0.50\textwidth]{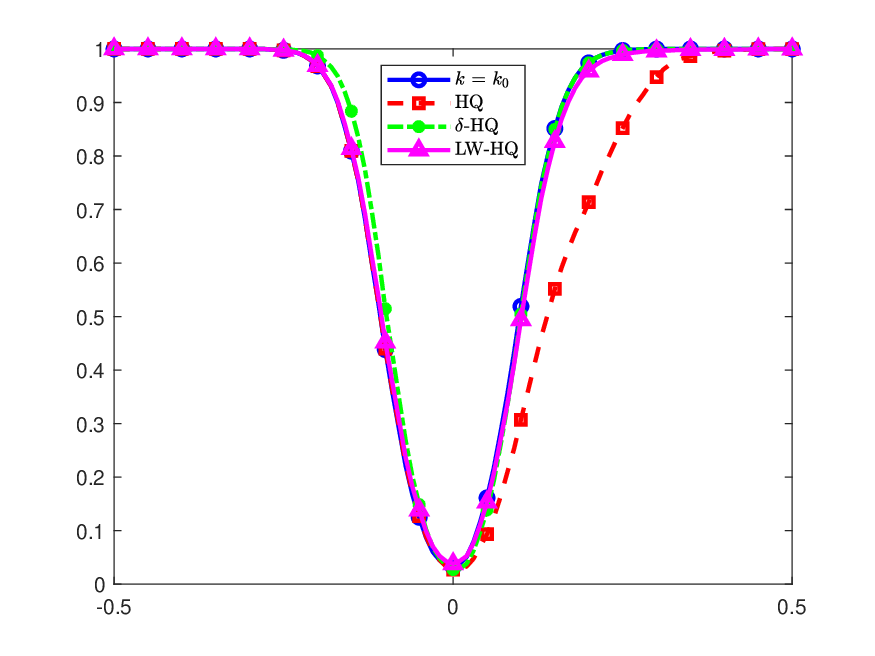}  
}
    \caption{ Simulated rejection frequencies of $H_0 : \delta = 0$, averaged over 5,000 replications. The DGP sets $k_0 = 1$ with $\beta_{0,0} = 0$ and $\beta_{1,0} = 1$, while $\epsilon_t \sim \text{NIID}(0,1)$ and $\delta_0 \in [-0.499, 0.499]$. The model is given by $y_t = \sum_{n = 0}^k \beta_n P_t (n) + \Delta^{-\delta} \epsilon_t$ with the order $k$ selected by either the usual time-domain HQ information criterion, the infeasible $\delta$-HQ criterion or our LW-HQ criterion.
    The correct specification with $k = k_0$ is included as a benchmark. The LM-statistic is based on the OLS residuals $\hat u_t (k)$ in \eqref{res}. For LW-HQ, the bandwidth is fixed at $m = \left\lfloor T^{0.65} \right\rfloor$.}%
    \label{f3}%
\end{figure}

Figure \ref{f3} illustrates simulated finite-sample power functions for the LM-type test. When the time-domain HQ criterion is used for order selection, the test exhibits a substantial loss in power for $\delta_0>0$, as predicted by Theorem \ref{l1}, particularly for smaller bandwidths $m$ and smaller sample sizes $T$. This power loss decreases as $m$ increases, but it remains visible even for comparatively large $m$. For the commonly used bandwidth $m=\lfloor T^{0.65}\rfloor$, the loss in power is especially pronounced. By comparison, the LM test using the true order ($k=k_0$) achieves reasonable power across all scenarios. The LM test based on the infeasible $\delta$-HQ also performs well and often tracks the benchmark closely. In some cases it even delivers higher rejection rates than the benchmark. However, as we show forthwith, this reflects systematic mis-selection of the polynomial order rather than a genuine improvement in performance. Importantly, our feasible LW-HQ yields power functions that are very close to those of the true-order benchmark.

\begin{landscape}
\begin{figure}[p] 
  \centering

\subfloat[$T=256,\ \delta_0=-0.3$]{\includegraphics[width=\fourcolw]{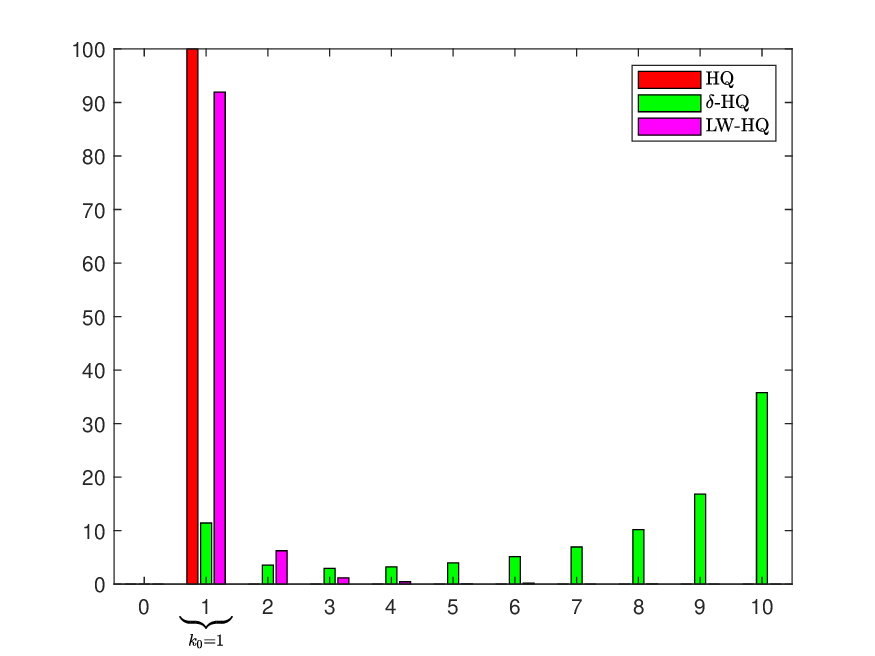}}
\subfloat[$T=256,\ \delta_0=0$]{\includegraphics[width=\fourcolw]{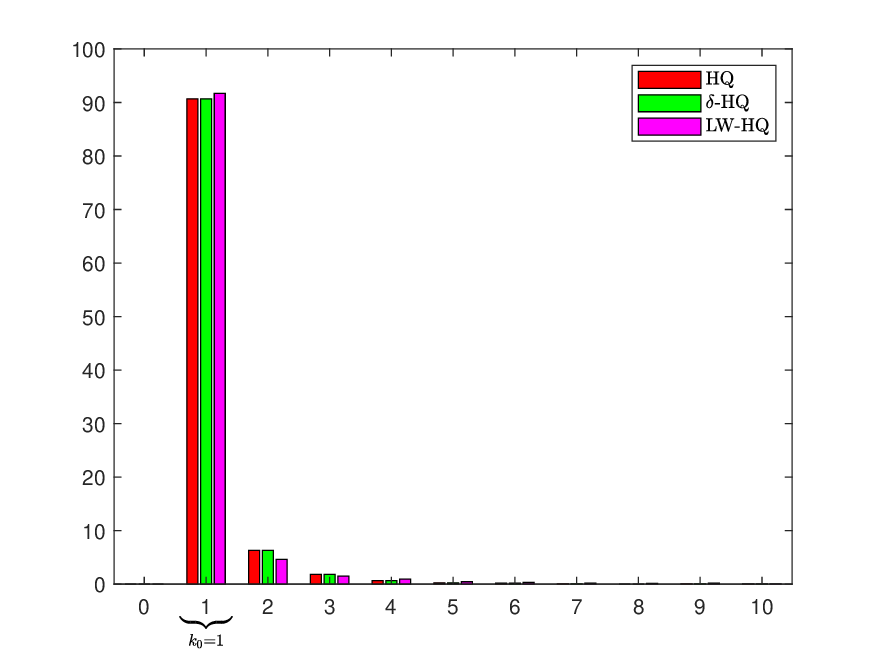}}
\subfloat[$T=256,\ \delta_0=0.3$]{\includegraphics[width=\fourcolw]{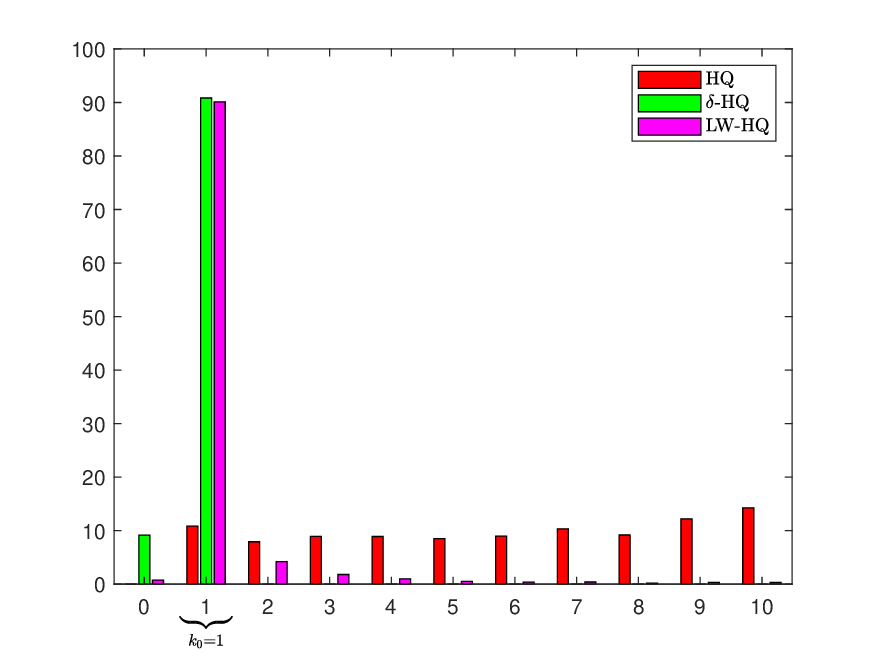}}
\subfloat[$T=256,\ \delta_0=0.45$]{\includegraphics[width=\fourcolw]{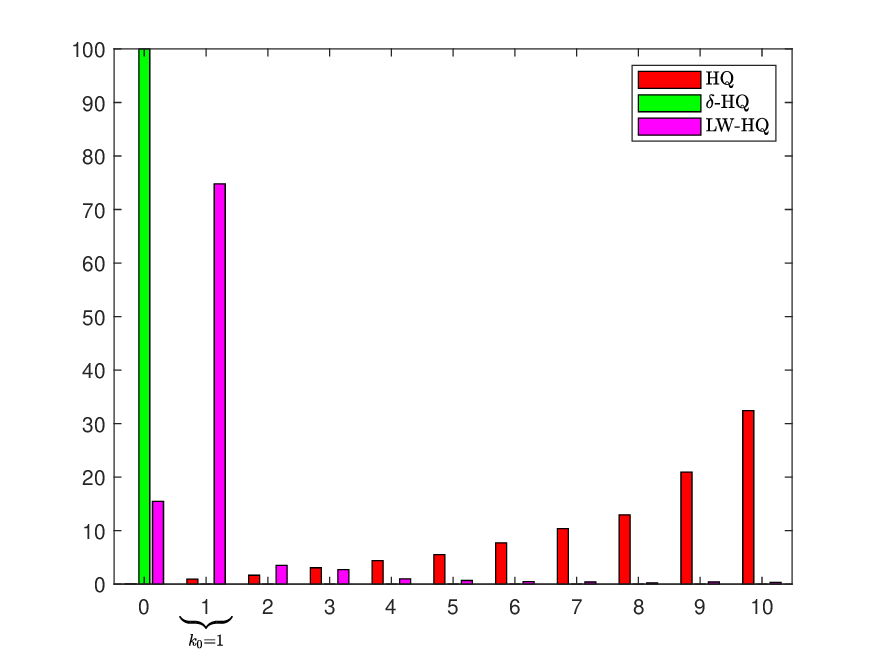}}

\medskip

\subfloat[$T=512,\ \delta_0=-0.3$]{\includegraphics[width=\fourcolw]{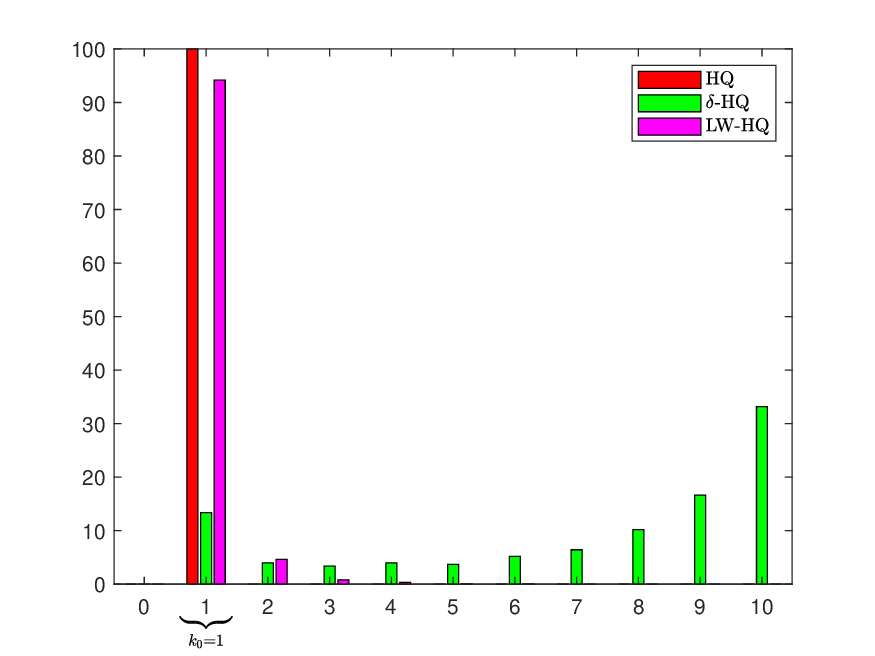}}
\subfloat[$T=512,\ \delta_0=0$]{\includegraphics[width=\fourcolw]{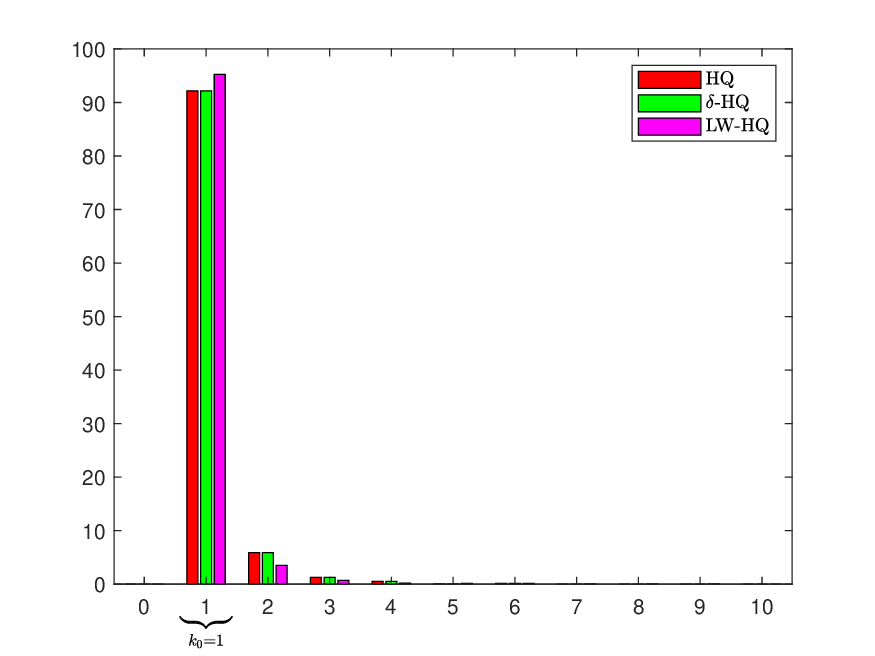}}
\subfloat[$T=512,\ \delta_0=0.3$]{\includegraphics[width=\fourcolw]{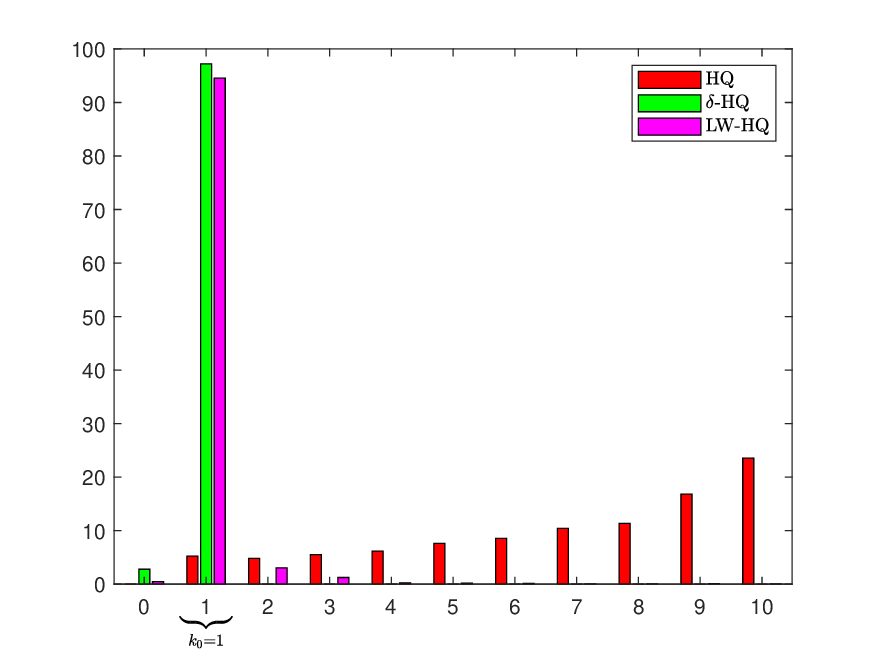}}
\subfloat[$T=512,\ \delta_0=0.45$]{\includegraphics[width=\fourcolw]{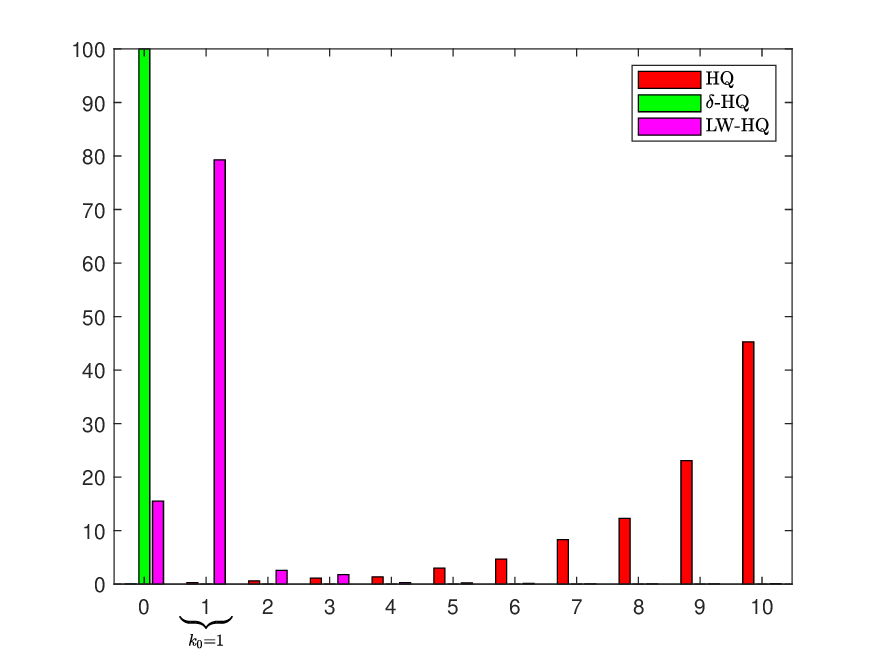}}

   \caption{Histograms of the estimated Chebyshev polynomial order selected by three HQ-type information criteria. The DGP is as in \eqref{eq1}--\eqref{eq21} with $u_t = \Delta^{-\delta_0} \eta_t$, where $\eta_t \sim \text{NIID}(0,1)$ and $\delta_0 \in \{-0.3, 0, 0.35, 0.45\}$. The true Chebyshev polynomial is specified by $k_0 = 1$, $\beta_{0,0} = 0$ and $\beta_{1,0} = 1$. The selection frequencies are averaged across 5,000 replications. The results for the time domain HQ criterion are shown in red, those for the infeasible $\delta$-HQ criterion proposed by \textcite{lavielle2000least}  in green, and for our LW-HQ criterion  in magenta. For LW-HQ we fix the bandwidth $m = \left\lfloor T^{0.65} \right\rfloor$.}
  \label{f4}
\end{figure}
\end{landscape}

Figure \ref{f4} displays histograms of the selected Chebyshev order for the DGP in \eqref{eq1}--\eqref{eq21}, where $\eta_t \sim \text{NIID}(0,1)$ and $\delta_0\in\{-0.3,0,0.35,0.45\}$. The true order of the Chebyshev polynomial is $k_0 = 1$, and the true coefficient values are set to $\beta_{0,0} = 0$ and $\beta_{1,0} = 1$ without loss of generality. The sample sizes considered are $T\in\{256,512\}$.
It turns out that the time-domain HQ criterion is accurate when $\delta_0\le 0$ (see panels (a),(b),(e),(f)), but for $\delta_0>0$ it tends to increasingly overselect large orders, in line with Theorem \ref{l1}. This upward bias also becomes more pronounced as $T$ increases (cf.\ panels (c)-(d), vs.\ (g)-(h)). The $\delta$-HQ criterion of \textcite{lavielle2000least} behaves reasonably at $\delta_0=0$ and $\delta_0= 0.3$, but under negative persistence ($\delta_0=-0.3$) it underpenalises and selects overly large $k$, whereas under strong persistence ($\delta_0=0.45$) it overpenalises and collapses to $k=0$ in the vast majority of replications. By contrast, our LW-HQ criterion concentrates its mass at the true order across all values of $\delta_0$ and both sample sizes, indicating stable order selection throughout.

Extensive simulations reported in the Supplementary Appendix (see Tables S.1-S.6 for size, Tables P.1-P.4 for power, and Tables O.1-O.6 for order selection) examine the performance of the LM test, the associated left- and right-tailed $t$-tests, and the order-selection procedures for a variety of other DGPs and model specifications: we allow the short-run component $\eta_t$ to follow IID, AR(1), and ARCH(1)-type processes, so as to better match features of empirical data. Throughout, the deterministic component is modelled by a Chebyshev polynomial with true order $k_0 \in \{0,1\}$ and varying magnitude of the corresponding coefficients. We let the bandwidth $m=\left\lfloor T^{\alpha}\right\rfloor$ vary with $\alpha \in \{0.40,\ldots,0.80\}$  and let the sample size $T \in \{256,512,1024\}$. In the following, we summarise the main findings.

First, the specification of the short-run dependence in $\eta_t$ matters for the finite sample performance of the LM test and for order selection. While the procedures are generally well behaved under IID and ARCH(1)-type specifications, the simulations show that AR(1)-type of dependence can lead to marked size distortions in the test, especially when the bandwidth is chosen too large. For empirically relevant one-sided alternatives, the right-tailed $t$-test behaves well yet can become oversized under $AR$-type short-run dependence. For order selection, the impact of short-run dependence is most visible for the time-domain criteria, which perform markedly worse in the AR(1) design than the LW-based procedures. LW-BIC, in particular, remains highly accurate once a moderate bandwidth is used.    

Second, our simulation results demonstrate that the bandwidth $m$ has a significant impact on the finite sample properties of the LM test and of the frequency-based order selection. Regarding the test, we recommend a bandwidth choice that is moderate, i.e.\ between
$m=\left\lfloor T^{0.50}\right\rfloor $ and $m=\left\lfloor T^{0.60}\right\rfloor$, which typically delivers the best size and power balance across the different designs. 
It is important to avoid larger bandwidths (e.g.\ between $m = \left\lfloor T^{0.65}\right\rfloor$ and $\left\lfloor T^{0.80}\right\rfloor$) as they can generate severe size distortions when the DGP contains short-run serial dependence beyond the fractional component. On the other hand, a very small $m$ is often conservative and can significantly reduce power. As for order selection, our additional simulations suggest that the probability of selecting the correct Chebyshev order under the LW-based criteria improves monotonically with $m$, while over-selection declines sharply as the bandwidth increases. In particular, values around $m=\lfloor T^{0.60}\rfloor$ already deliver highly accurate order selection and, at the same time, satisfy the bandwidth condition in Assumption \ref{bandw}. The LW-BIC performs best, with LW-HQ closely behind. The time-domain BIC or HQ criteria should be avoided because they are not consistent under long memory. Although the $\delta$-BIC and $\delta$-HQ criteria also perform well in some cases, they are infeasible in practice and therefore are best viewed as benchmark procedures.

Overall, we recommend that either the LM-test or the $t$-test be combined with LW-BIC and that a moderate bandwidth be used in practice. In particular, $m=\lfloor T^{0.60}\rfloor$ provides a satisfactory overall compromise across the designs considered in our simulations, performing well in terms of both Chebyshev order selection and finite-sample properties of the tests, while also satisfying the bandwidth condition in Assumption \ref{bandw}. Accordingly, in the empirical application in Section \ref{s4}, we implement LW-BIC with $m=\lfloor T^{0.60}\rfloor$.

\section{Empirical illustration: UK Great Rations}\label{s4}

\textcite{kaldor1961capital} presents a series of ``stylized facts'' delineating constant relationships among certain macroeconomic variables. \textcite{klein1961some} essentially refer to them as Great Ratios. The concept of Great Ratios has given rise to a considerable empirical research body, which includes recent studies by \textcite{kapetanios2020time} and \textcite{chudik2023revisiting}. Of particular interest has been the question of whether the Great Ratios follow integrated processes. 
\textcite{kapetanios2020time} show that when analysing several ratios computed with UK data they display $I(1)$ behaviour although they can be described as $I(0)$ when a slowly changing deterministic component is included in the model. We broaden this debate by considering fractional alternatives. \textcite{kapetanios2020time} base their testing approach on the KPSS statistic. A KPSS-type test is, however, inappropriate in our setting since \textcite{shao2007local} establish that it has limited power against local fractional alternatives, namely of order $\log(T)^{-1}$. As opposed to that, our tests  possess substantial power against local fractional alternatives, as shown in the Monte Carlos simulations in Section \ref{s3}.

\begin{figure}[H]
  \centering
  \subfloat[$\log(C/Y)$]{
    \includegraphics[width=0.33\textwidth]{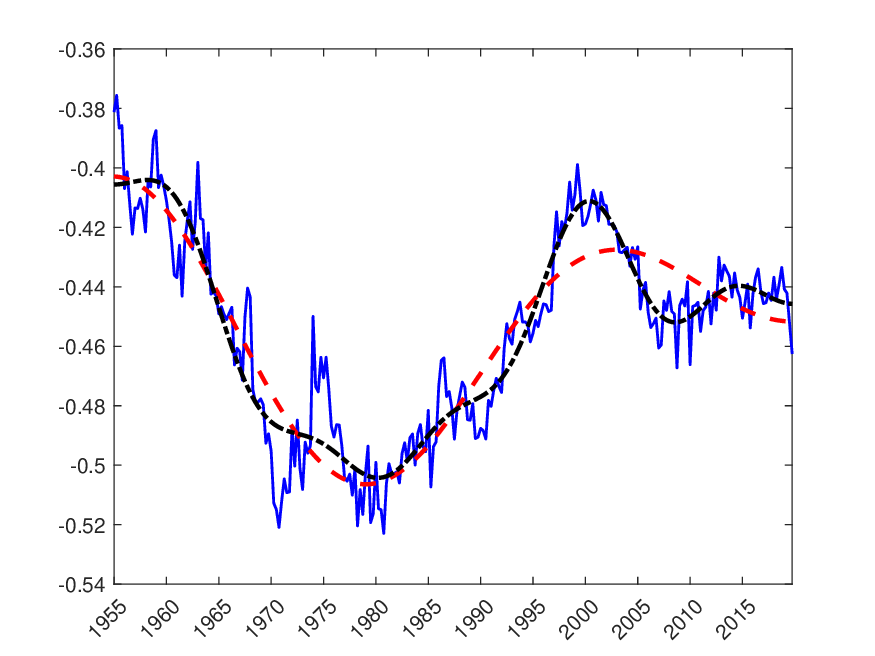}
  }
  \subfloat[$\log(I/Y)$]{
    \includegraphics[width=0.33\textwidth]{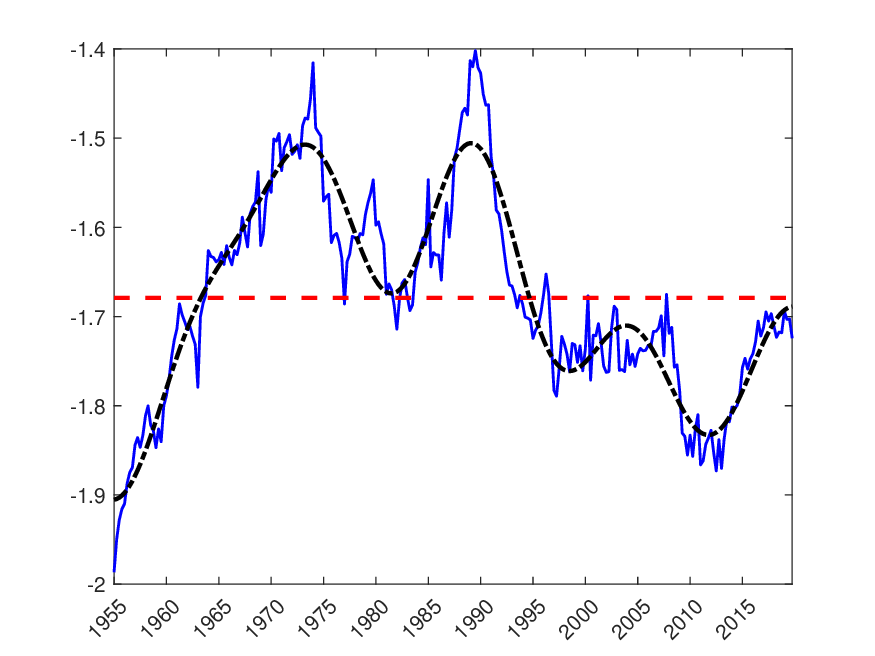}
  } 
    \subfloat[$\log(G/Y)$]{
    \includegraphics[width=0.33\textwidth]{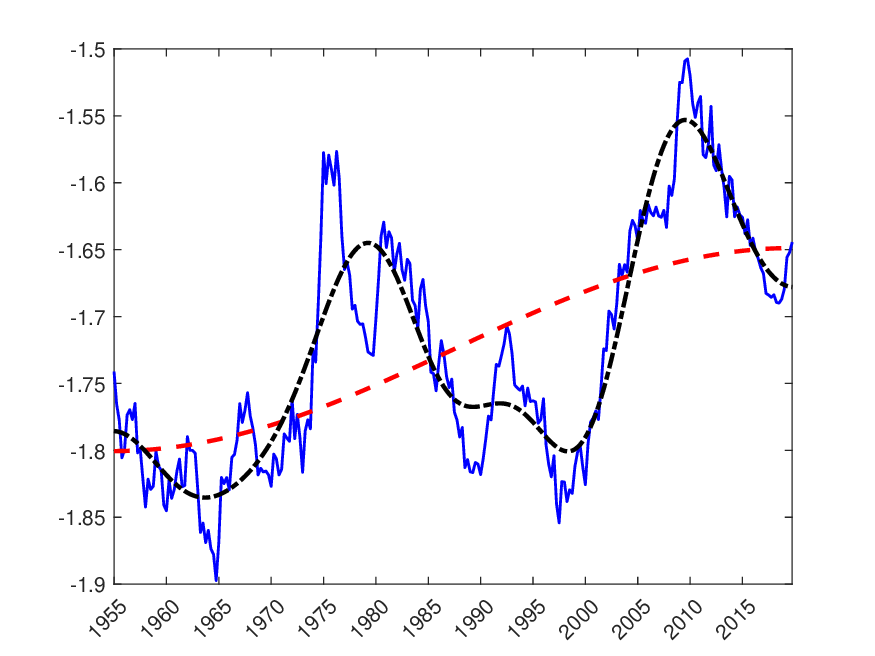}
  } \\
    \subfloat[$\log(X/Y)$]{
    \includegraphics[width=0.33\textwidth]{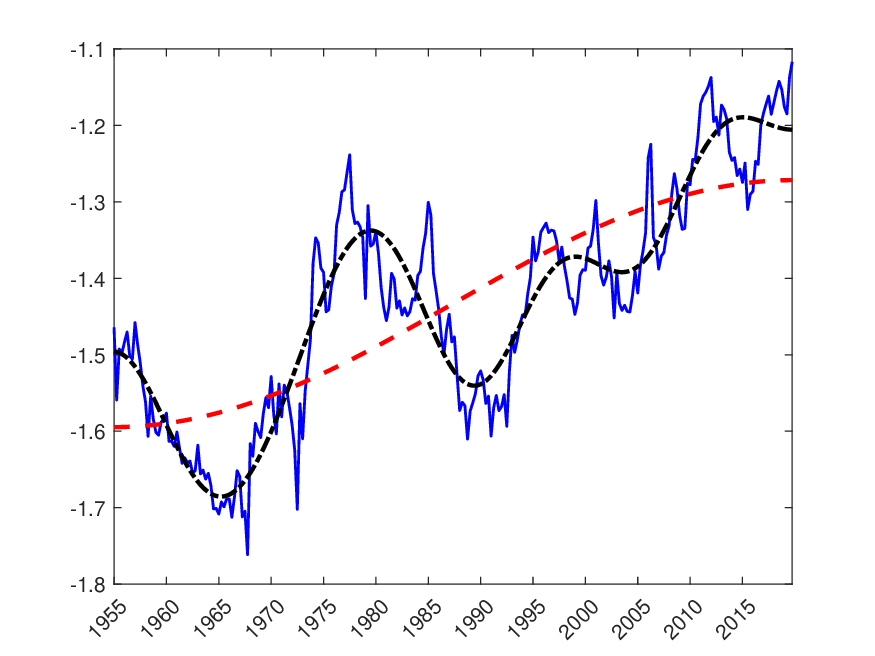}
    } 
    \subfloat[$\log(M/Y)$]{
    \includegraphics[width=0.33\textwidth]{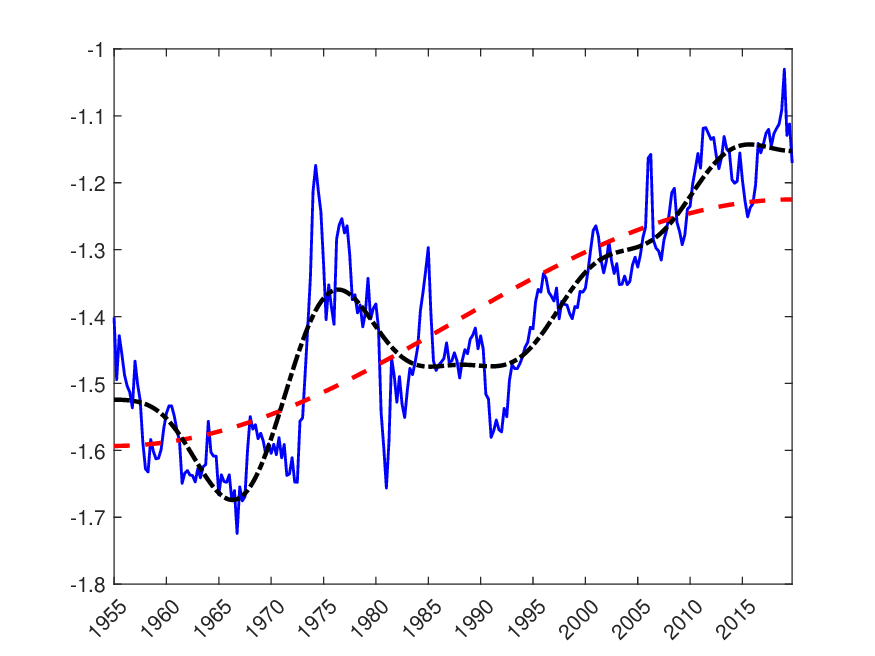}
  }
    \subfloat[$\log(\text{Labour share})$]{
    \includegraphics[width=0.33\textwidth]{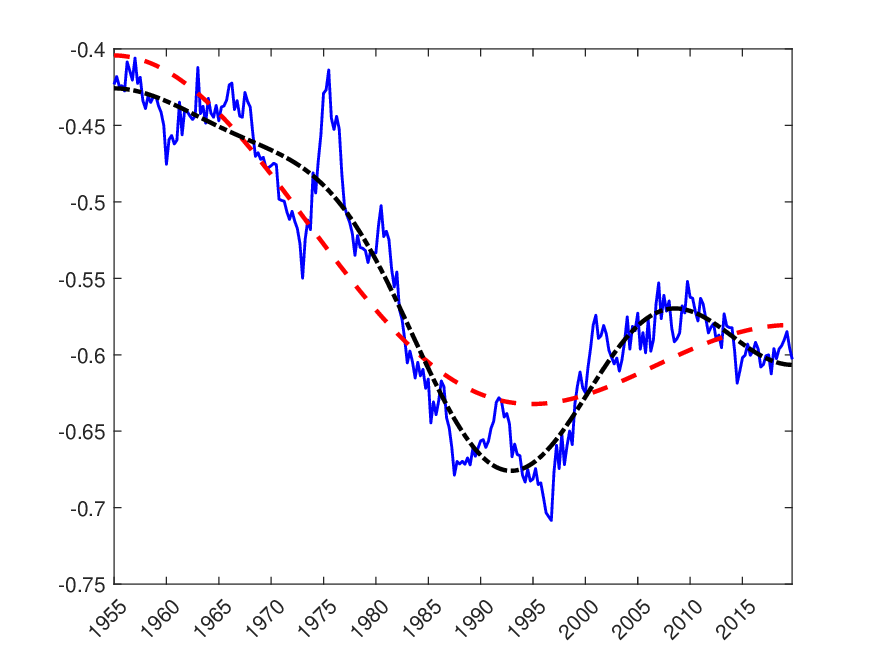}
    } \\
    \subfloat[$\log(\text{Profit share})$]{
    \includegraphics[width=0.33\textwidth]{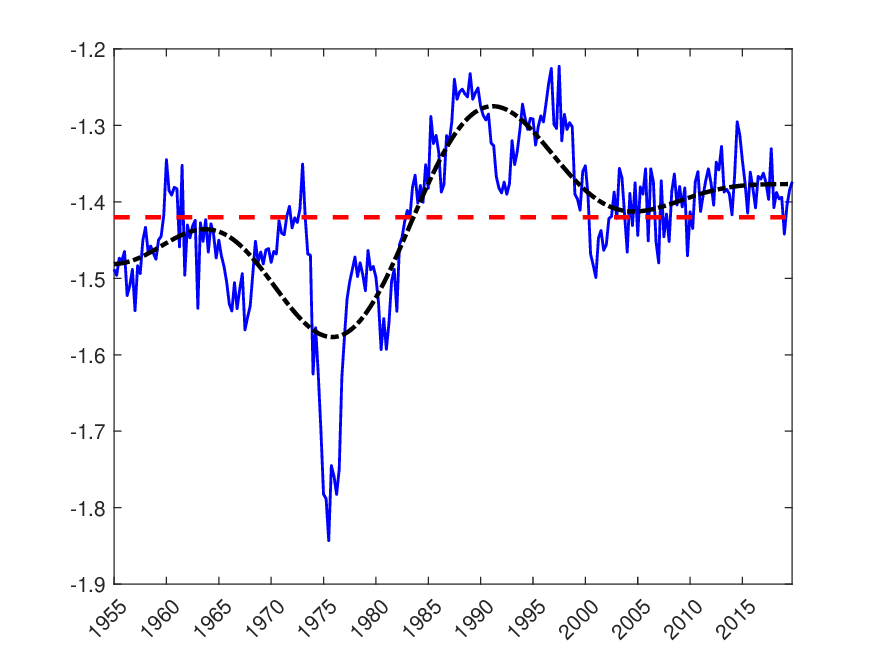}
    }
   \caption{ Logged UK Great Ratios. Dotted lines show the estimated deterministic component $\sum_{n=0}^{\hat{k}}\hat{\beta}_n P_t(n)$ with the order $\hat{k}$ selected by time domain BIC (black) and frequency-domain LW-BIC (red).} 
  \label{GR}%
\end{figure}

We construct the nominal Great Ratios using quarterly UK data from 1955Q1 to 2019Q4, i.e.\ up until the onset of the COVID-19 pandemic, yielding $T=260$ observations. To facilitate direct comparison with \textcite{kapetanios2020time}, we also consider the subsample 1955Q1--2015Q4. Variable definitions follow \textcite[p.\ 34]{kapetanios2019time}. All ratios are in logs. As recommended by our simulations in Section \ref{s3b}, we use the LW-BIC criterion for selecting the order of the Chebyshev polynomial. Both our LM-type statistic and our LW-BIC are computed using a bandwidth of $m=\lfloor T^{0.6}\rfloor$.  We set the maximal Chebyshev polynomial order equal to $K = 10$ and restrict the fractional parameter to $\delta\in[\Delta_1,\Delta_2]$ with $\Delta_1=-0.499$ and $\Delta_2 = 0.499$. Figure \ref{GR} plots the data series with the estimated deterministic component $\sum_{n=0}^{\hat{k}}\hat\beta_n P_t(n)$ superimposed: time-domain BIC selections are shown in black and frequency-domain LW-BIC in red. The $\delta$-BIC of \textcite{lavielle2000least} is infeasible. Under time-domain BIC, the fitted component follow the data closely because the selected polynomial orders are near the upper bound, producing plots that resemble those in \textcite{kapetanios2020time}, see also the values of $\hat k$ in Panel A of Table \ref{emp}. On the other hand, LW-BIC selects parsimonious polynomial orders, yielding fewer Chebyshev functions, cf.\ $\hat k$ in Panel B of Table \ref{emp}. The table also reports $t_{\hat{u}(\hat{k})}(\delta_0;m)$ for testing $H_0:\delta=0$ against $H_1:\delta>0$. With $k$ selected by the time-domain BIC, the $t$-statistics are small (i.e.\ less than 1.66) in both the full sample (1955Q1–2019Q4) and the subsample (1955Q1–2015Q4), so the $t$-test cannot reject $H_0$ at conventional levels. This reflects the tendency of time-domain BIC to overestimate $k$ and is consistent with Theorem \ref{l1}. Using LW-BIC for the selection of $k$, however, we reject $H_0$ at the 1\% level for all seven ratios in both samples, indicating pronounced long memory. Our findings therefore differ from those of \textcite{kapetanios2020time}, who conclude that the nominal Great Ratios are $I(0)$ once slowly varying deterministic components are allowed: Our fractional test continues to reject $H_0 : \delta=0$ at the 1\% level for all seven ratios, even when we use the larger sample. This effectively implies that the macroeconomic time series can be seen as fractionally cointegrated.

\begin{table}[H]
\centering
\caption{Tests of $H_0\!: \delta=0$ vs.\ $H_1\!: \delta>0$ for the Great Ratios when the order of the Chebyshev polynomial is selected by the time-domain BIC and our frequency-domain LW-BIC. The variables are: $C$ = consumption, $Y$ = output, $I$ = investment, $G$ = government expenditure, $X$ = exports, $M$ = imports, labour share, and profit share. *** denote significance at the 1\% level (one-sided $t$-test).}
\label{emp}
\setlength{\tabcolsep}{3pt}
\renewcommand{\arraystretch}{1.05}
\tiny
\begin{tabularx}{\linewidth}{ll*{7}{Y}}
\toprule
\multicolumn{9}{l}{\textbf{Panel A: Time-domain BIC selection}}\\
\midrule
Year & Statistic
& $\log(C/Y)$ & $\log(I/Y)$ & $\log(G/Y)$ & $\log(X/Y)$ & $\log(M/Y)$
& $\log(\text{Lab.\ share})$ & $\log(\text{Prof.\ share})$ \\
\midrule
1955Q1--2019Q4 & $t_{\hat{u}(\hat{k})}(\delta_0;m)$
& 0.32 & 0.57 & 0.80 & 0.34 & 0.19 & 1.66 & 1.29 \\
 & $\hat{k}$
& 10 & 10 & 10 & 9 & 10 & 5 & 6 \\
\addlinespace[0.2em]
1955Q1--2015Q4 & $t_{\hat{u}(\hat{k})}(\delta_0;m)$
& 0.52 & 0.66 & 1.02 & 0.13 & 0.19 & 0.86 & 0.89 \\
 & $\hat{k}$
& 9 & 9 & 9 & 10 & 10 & 10 & 10 \\
\midrule
\multicolumn{9}{l}{\textbf{Panel B: Frequency-domain LW-BIC selection}}\\
\midrule
Year & Statistic
& $\log(C/Y)$ & $\log(I/Y)$ & $\log(G/Y)$ & $\log(X/Y)$ & $\log(M/Y)$
& $\log(\text{Lab.\ share})$ & $\log(\text{Prof.\ share})$ \\
\midrule
1955Q1--2019Q4 & $t_{\hat{u}(\hat{k})}(\delta_0;m)$
& 3.06*** & 9.49*** & 7.19*** & 6.53*** & 5.55*** & 6.40*** & 7.54*** \\
 & $\hat{k}$
& 3 & 0 & 1 & 1 & 1 & 2 & 0 \\
\addlinespace[0.2em]
1955Q1--2015Q4 & $t_{\hat{u}(\hat{k})}(\delta_0;m)$
& 3.06*** & 9.70*** & 7.43*** & 6.32*** & 5.30*** & 6.66*** & 7.42*** \\
 & $\hat{k}$
& 3 & 0 & 1 & 1 & 1 & 2 & 0 \\
\bottomrule
\end{tabularx}
\end{table}

\section{Conclusion}\label{s5}

This paper addresses the challenge of disentangling smooth  deterministic trends from the order of fractional integration. Building on the frameworks proposed by \textcite{cuestas2016testing} and \textcite{iacone2022semiparametric}, we develop a semi-parametric testing procedure for the order of fractional integration that models smooth trends using a Chebyshev polynomial. A local Whittle approximation is employed to mitigate issues arising from unmodelled short-range dependence. A key contribution is a novel frequency-domain information criterion that consistently selects the polynomial order, even under long memory. The criterion measures model fit using the local Whittle objective computed from OLS residuals and controls for model complexity through a penalty that depends only on the bandwidth $m$.

We prove the asymptotic validity of the proposed testing and selection procedures and our simulation study confirms their favourable finite-sample performance. Our empirical application to the UK Great Ratios provides new evidence suggesting the presence of fractional cointegration among these important macroeconomic variables.

Beyond our specific setting, the frequency-based information criterion we propose has broader potential. First, it could be employed to address similar model selection challenges in the framework of \textcite{iacone2022semiparametric}, where determining the number of structural breaks is complicated by the presence of long memory and where standard time-domain information criteria select too many breaks when $\delta_0 > 0$. 
Using our local Whittle information criterion could mitigate this tendency. Secondly, the same approach could be adapted to alternative orthogonal representations of deterministic smooth trends, such as those considered by \textcite{perron2020trigonometric} and \textcite{abadir2011d}.

\printbibliography[title={References}]

\appendix
\counterwithin*{equation}{section} 
\renewcommand\theequation{\thesection.\arabic{equation}} 

\section{Appendix}

This appendix provides supporting technical arguments for the main results. Lemmas \ref{consCheby} and \ref{lemma:w_P} establish rates for the OLS coefficients and the behavior of the periodogram of Chebyshev functions. Lemma \ref{lemma:wtimesu} derives bounds for weighted cross-products between $u_t$ and Chebyshev function $P_t(s)$ in the frequency domain, and Lemma \ref{thm6} establishes consistency and asymptotic normality of the local Whittle estimator based on fitted residuals. The remaining subsections provide the proofs of Theorems \ref{thm1}, \ref{thm2}, \ref{l1}, \ref{thm3}, and \ref{thm5}.

\subsection{Preliminary results}

\begin{lemma}\label{consCheby}
Let $y_t$ be generated according to \eqref{eq1}--\eqref{eq21} and let Assumption \ref{ass2} hold. Under $H_c$, for any fixed $k$,
\begin{align*}
     T^{1/2-\delta_0} \left(\hat\beta(k)-\beta_0(k)\right) = O_p(1),
\end{align*}
for $\delta_0 \in (-0.5,0.5)$, where $\hat\beta(k)=(X(k)'X(k))^{-1}X(k)'y$ is the OLS estimator based on the regressors $(P_t(0),\ldots,P_t(k))$, and
\begin{align*}
\beta_0(k) =
\bigl(\beta_{0,0},\beta_{1,0},\ldots,\beta_{k,0}\bigr)' \in \mathbb{R}^{k+1},
\qquad
\beta_{n,0} =0 \ \text{for } n>k_0 .
\end{align*}
Furthermore,
\begin{align*}
    \Var\left( T^{1/2-\delta_0} \hat\beta(k) \right) = O(1),
\end{align*}
under $H_0$. 
\end{lemma}
\begin{proof}
For any non-negative  $m, n$
\begin{align}
\frac{1}{T}\sum_{t=1}^T P_t(m)P_t(n)=
\begin{cases}
2, & m=n=0,\\
1, & m=n>0,\\
0, & m\neq n,
\end{cases} \label{crossp}
\end{align}
see \textcite[eq. (6.12)]{mason2002chebyshev}. Hence
\begin{align}
    T^{-1}X(k)'X(k) &= D(k) \label{orth}
\end{align}
where $D(k) = \mathrm{diag}(2,1,\ldots,1)$. 

Note
\begin{align*}
T^{1/2-\delta_0} \left(\hat\beta(k)-\beta_0(k)\right) =\left( T^{-1}X(k)'X(k)\right)^{-1} T^{1/2-\delta_0} T^{-1}X(k)'u.
\end{align*}
Therefore, using \eqref{orth},
\begin{align*}
T^{1/2-\delta_0} \left(\hat\beta(k)-\beta_0(k)\right) = D(k)^{-1} T^{-1/2-\delta_0} X(k)'u.
\end{align*}
Since $D(k)$ is deterministic and non-singular for any fixed $k$, it suffices to show 
\begin{align*}
    T^{-1/2-\delta_0} X(k)'u = O_p(1) 
\end{align*}
or equivalently, for each $s \leq k$,
\begin{align}
    T^{-1/2-\delta_0} \sum_{t = 1}^T P_t(s) u_t  = O_p(1). \label{crossup}
\end{align}

Define the normalised partial-sum process
\begin{align*}
S(\tau)=T^{-1/2-\delta_0}\sum_{t=1}^{\lfloor \tau T\rfloor}u_t.
\end{align*}
for $\tau \in [0,1]$.
Under $H_c$ and Assumption \ref{ass2}, $S(\cdot)\Rightarrow\kappa(\delta_0)\,W(\cdot,\delta_0)$, where $W(\cdot,\delta_0)$ is a type-I fractional Brownian motion, see \textcite[Lemma 1]{iacone2022semiparametric}. Since $W(\cdot,\delta_0)$ has almost surely continuous sample paths on $[0,1]$, application of the continuous mapping theorem yields
\begin{align*}
\sup_{\tau \in [0,1]} |S(\tau)| = O_p(1).     
\end{align*}
In particular,
\begin{align*}
    \max_{1\leq t \leq T}|S(t/T)| \leq  \sup_{\tau \in [0,1]}|S(\tau)|  = O_p(1).
\end{align*}

Fix $s \leq k$. Summation by parts yields
\begin{align*}
     T^{-1/2-\delta_0} \sum_{t = 1}^T P_t(s) u_t  &= P_T(s) T^{-1/2-\delta_0} \sum_{t = 1}^T u_t - \sum_{t = 1}^{T-1} \left( P_{t+1}(s) - P_t(s)  \right) T^{-1/2-\delta_0}  \sum_{j = 1}^{t} u_t  \\
     &= P_T(s) S(1) - \sum_{t = 1}^{T-1} \left( P_{t+1}(s) - P_t(s)  \right)  S(t/T) . 
\end{align*}

\noindent\textbf{Case $s=0$.}
Since $P_t(0) = \sqrt{2}$ for all $t$, we have $P_{t+1}(0)-P_t(0) = 0$ and thus 
\begin{align*}
     T^{-1/2-\delta_0} \sum_{t = 1}^T P_t(0) u_t = P_T(0) S(1) \leq P_T(0) \max_{1\leq t \leq T} |S(t/T)| = O_p(1).
\end{align*}

\noindent\textbf{Case $s>0$.}
For $s>0$, using $\cos(a)-\cos(b) = -2\sin((a+b)/2)\sin((a-b)/2)$ (  \textcite[eq. 4.3.37]{abramowitz1968handbook}) with $a = s \pi (t+1/2)/T$ and $b = s \pi (t-1/2)/T$, yields
\begin{align*}
   P_{t+1}(s)-P_{t}(s) &= \sqrt{2} \cos \left(s \pi \frac{t+1/2}{T} \right) - \sqrt{2} \cos \left(s \pi \frac{t - 1/2}{T} \right) \\
   &= -2  \sqrt{2} \sin\left(\frac{s\pi}{2T}\right) \sin\left(\frac{s \pi t}{T}\right).
\end{align*}
Hence, since $|\sin(x)| \leq x$ for $x\geq 0$,
\begin{align*}
   |P_{t+1}(s)-P_{t}(s)|
   &= 2  \sqrt{2} \left|\sin\left(\frac{s\pi}{2T}\right)\right| \left|\sin\left(\frac{s \pi t}{T}\right)\right| \\
   &= C \frac{1}{T}.
\end{align*}
Taking the absolute value and using the bound above
\begin{align*}
  |P_T(s)  S(1) - \sum_{t = 1}^{T-1} \left( P_{t+1}(s) - P_{t}(s) \right) S(t/T)| &\leq |P_T(s)|  |S(1)| \\ 
  &\ \ \ +    \sum_{t = 1}^{T-1}  |\left( P_{t+1}(s) - P_{t}(s) \right)| | S(t/T)| \\
  &\leq  |P_T(s)|  |S(1)| + \frac{C}{T}  \sum_{t = 1}^{T-1} | S(t/T)|  \\
  &\leq  |P_T(s)|  |S(1)| +   \frac{C}{T} T  \max_{1\leq t \leq T} |S(t/T)| \\
  &\leq  |P_T(s)|  \max_{1\leq t \leq T} |S(t/T)| +   C  \max_{1\leq t \leq T} |S(t/T)|.
\end{align*}
Finally, $|P_T(s)| \leq C$ for all $s$. Hence, 
\begin{align*}
    \left| T^{-1/2-\delta_0} \sum_{t = 1}^T P_t(s) u_t \right|  &\leq C \left(|S(1)| + \max_{1\leq t \leq T} |S(t/T)| \right) = O_p(1).
\end{align*}
Therefore $T^{-1/2-\delta_0} X(k)'u = O_p(1)$ and hence
\begin{align*}
    T^{1/2-\delta_0} \left(\hat\beta(k)-\beta_0(k)\right) = D(k)^{-1} T^{-1/2-\delta_0} X(k)'u = O_p(1).
\end{align*}

It remains to show $\Var(T^{1/2-\delta_0} \hat{\beta}(k)) = O(1)$. 
Since 
\begin{align*}
    T^{1/2-\delta_0} \hat{\beta}(k) = T^{1/2-\delta_0} \beta_0(k) + D(k)^{-1} T^{-1/2-\delta_0} X(k)'u, 
\end{align*}
and the first term is non-random, it suffices to show
\begin{align*}
    \Var\left( T^{-1/2-\delta_0} X(k)'u  \right) = O(1).
\end{align*}
By Cauchy-Schwarz, it is enough to bound the diagonal elemnts, i.e.\ for each $s \leq k$,
\begin{align*}
      \Var\left( T^{-1/2-\delta_0} \sum_{t = 1}^T P_t(s) u_t  \right) = O(1).
\end{align*}
From the summation by part representation above, $|P_t(s)| \leq C$ and $|P_{t+1}(s)-P_{t}(s)| \leq C/T$, and using $(a+b)^2 \leq 2 a^2 + 2b^2$ together with Jensen's inequality, we obtain
\begin{align*}
     E \left(T^{-1/2-\delta_0} \sum_{t = 1}^T P_t(s) u_t \right)^2 \leq  C E S(1)^2 + C \frac{1}{T} \sum_{t = 1}^{T-1}   E S(t/T)^2
\end{align*}
Hence, it suffices to show $E S(\tau)^2 \leq C$ uniformly for $\tau \in [0,1]$. Let $n = \lfloor \tau T\rfloor$. Then 
\begin{align*}
    E S(\tau)^2 = \Var \left( T^{-1/2-\delta_0} \sum_{t = 1}^n  u_t \right) = T^{-1-2\delta_0} \Var \left( \sum_{t = 1}^n  u_t \right)
\end{align*}
Under $H_0$ and Assumption \ref{ass2}, \textcite{abadir2009two}, eq. (1.2), implies 
\begin{align*}
    \Var \left( n^{-1/2-\delta_0} \sum_{t = 1}^n  u_t \right) = O(1),
\end{align*}
so 
\begin{align*}
    \Var \left( \sum_{t = 1}^n  u_t \right) = O( n^{1+2\delta_0}).
\end{align*}
Therefore 
\begin{align*}
    E S(\tau)^2 \leq C T^{-1-2\delta_0} n^{1+2\delta_0} \leq C \left( \frac{n}{T} \right)^{1+2\delta_0} \leq C,
\end{align*}
since $n = \lfloor \tau T\rfloor \leq T$.

\end{proof}

\begin{lemma}\label{lemma:w_P}
    Let $m \rightarrow \infty$ and $m = o(T)$. Then for any integer $s \geq 1$ and uniformly for $1\leq j \leq m$,
    \begin{align*}
        I_{P(s)}(\lambda_j) = T g_{j,s}(1 + o(1)),
    \end{align*}
    where $g_{j,s} > 0$ for all $j$ when $s$ is odd, while when $s$ is even $g_{j,s} > 0$ for $j = s/2$ and $g_{j,s} = 0$ for $j \neq s/2$. Moreover, $g_{j,s} = O(j^{-2})$ for all $s\geq 1$ as $j \rightarrow \infty$. 
\end{lemma}
\begin{proof}
    Using $\cos(x) = (e^{ix} + e^{-ix})/2$ (\textcite[eq. 4.3.2]{abramowitz1968handbook}) yields
    \begin{align*}
        P_t(s) &= \frac{\sqrt{2}}{2} \left(e^{i\frac{s\pi}{T}(t-\frac{1}{2})} + e^{-i\frac{s\pi}{T}(t-\frac{1}{2})} \right)\\
        &= \frac{\sqrt{2}}{2} \left( e^{-i\frac{s\pi}{T}\frac{1}{2}}  e^{i\frac{s\pi}{T}t} + e^{i\frac{s\pi}{T}\frac{1}{2}}  e^{-i\frac{s\pi}{T}t} \right)
    \end{align*}
    Therefore,    
    \begin{align}
        w_{P(s)}(\lambda_j) &= \frac{1}{2\sqrt{\pi T}} \sum_{t = 1}^T \left( e^{-i\frac{s\pi}{2T}}  e^{i(\frac{s\pi}{T} + \lambda_j )t} + e^{i\frac{s\pi}{2T}}  e^{i(\lambda_j - \frac{s\pi}{T} )t} \right) \nonumber \\
        &= \frac{1}{2\sqrt{\pi T}}  e^{-i\frac{s\pi}{2T}} \sum_{t = 1}^T e^{i(\frac{s\pi}{T} + \lambda_j )t} + \frac{1}{2\sqrt{\pi T}}  e^{i\frac{s\pi}{2T}} \sum_{t = 1}^T  e^{i(\lambda_j - \frac{s\pi}{T} )t} \label{w_P}
    \end{align}

    For any real $\alpha$,
    \begin{align*}
        \sum_{t = 1}^T e^{it\alpha} = e^{i \alpha} \frac{1-e^{i T\alpha}}{1-e^{i \alpha}},
    \end{align*}
    see \textcite[eq. 3.1.10]{abramowitz1968handbook}. Also
    \begin{align*}
        \sin(z) = \frac{e^{iz}-e^{-iz}}{2i}
    \end{align*}
    see \textcite[eq. 4.3.1]{abramowitz1968handbook} and rewriting yields
     \begin{align*}
       1-e^{i2z} = -e^{iz} 2i \sin(z).
    \end{align*}
    Hence
      \begin{align*}
        \sum_{t = 1}^T e^{it\alpha} = e^{i \alpha (T+1)/2} \frac{\sin(T\alpha/2)}{\sin(\alpha/2)}.
     \end{align*}
Substituting this expression into $ w_{P(s)}(\lambda_j)$ yields
\begin{align*}
    w_{P(s)}(\lambda_j) &= \frac{1}{2\sqrt{\pi T}}  e^{-i\frac{s\pi}{2T}} e^{i (\frac{s\pi}{T} + \lambda_j ) (T+1)/2} \frac{\sin(T(\frac{s\pi}{T} + \lambda_j )/2)}{\sin((\frac{s\pi}{T} + \lambda_j )/2)} \\
    \ \ \ &+ \frac{1}{2\sqrt{\pi T}}  e^{i\frac{s\pi}{2T}}  e^{i (\lambda_j - \frac{s\pi}{T} ) (T+1)/2} \frac{\sin(T(\lambda_j - \frac{s\pi}{T} )/2)}{\sin((\lambda_j - \frac{s\pi}{T} )/2)}
\end{align*}
Since $\lambda_j = 2\pi j/T$, we have  $s\pi/T + \lambda_j = \pi (2 j + s)/T$ and $\lambda_j - s\pi/T  = \pi (2 j-s)/T$. Moreover, $ e^{-i s\pi/(2T)} e^{i (s\pi/T + \lambda_j ) (T+1)/2}  = e^{i \lambda_j (T+1)/2}  e^{-i s\pi/(2T) +  i s\pi (T+1) /(2T)  } = e^{i \lambda_j (T+1)/2}  e^{i s\pi/2  }$ and similarly $e^{i s\pi/(2T)} e^{i ( \lambda_j - s\pi/T ) (T+1)/2} =  e^{i \lambda_j (T+1)/2}  e^{-i s\pi/2  }$. Hence 
\begin{align*}
    w_{P(s)}(\lambda_j) &= \frac{1}{2\sqrt{\pi T}}  e^{i \lambda_j \frac{ (T+1)}{2}} \left( e^{i \frac{s\pi}{2}  } \frac{\sin(\pi (2 j + s)/2)}{\sin(\pi (2 j + s)/(2T))} + e^{-i \frac{s\pi}{2}  } \frac{\sin(\pi (2 j - s)/2)}{\sin(\pi (2 j - s)/(2T))} \right) 
\end{align*}
Since $|e^{i \lambda_j \frac{ (T+1)}{2}} | = 1$ 
\begin{align*}
    I_{P(s)}(\lambda_j) = \frac{1}{4 \pi T}  \left| e^{i \frac{s\pi}{2}  } \frac{\sin(\pi (2 j + s)/2)}{\sin(\pi (2 j + s)/(2T))} + e^{-i \frac{s\pi}{2}  } \frac{\sin(\pi (2 j - s)/2)}{\sin(\pi (2 j - s)/(2T))} \right|^2 
\end{align*}
 For any real $a$ and $b$
\begin{align*}
    \left| e^{i \frac{s\pi}{2}  } a + e^{-i \frac{s\pi}{2}  } b \right|^2  = a^2 + b^2 + 2 ( e^{-i s\pi  }  +  e^{i s\pi  } ) a b  = a^2 + b^2 + 2 \cos(s\pi) a b = (b+ (-1)^s a)^2
\end{align*}
using $\cos(x) = (e^{ix} + e^{-ix})/2$ and $\cos(s\pi) = (-1)^s$.
Therefore 
\begin{align*}
    I_{P(s)}(\lambda_j) = \frac{1}{4\pi T}  \left( \frac{\sin(\pi (2 j - s)/2)}{\sin(\pi (2 j - s)/(2T))} +  (-1)^s \frac{\sin(\pi (2 j + s)/2)}{\sin(\pi (2 j + s)/(2T))} \right)^2. 
\end{align*}

\noindent\textbf{Case $s$ is odd.} Assume $s$ is odd and $1 \leq j \leq m$ with $m = o(T)$. Then $(2j \pm s)/(2T) \rightarrow 0$ and the expansion $\sin(x) = x + O(x^3)$ as $x \rightarrow 0$ gives
\begin{align*}
    I_{P(s)}(\lambda_j) &= \frac{1}{4\pi T} \left( \frac{2T}{\pi} \right)^2  \left( \frac{\sin(\pi (2 j - s)/2)}{2 j - s}  +  (-1)^s \frac{\sin(\pi (2 j + s)/2)}{2 j + s} \right)^2 (1+o(1))  \\
    &= \frac{T}{\pi^3}  \left( \frac{\sin(\pi (2 j - s)/2)}{2 j - s} - \frac{\sin(\pi (2 j + s)/2)}{2 j + s} \right)^2 (1+o(1)).
\end{align*}
Hence $I_{P(s)}(\lambda_j) = T g_{j,s} (1+o(1))$.

Moreover, when $s$ is odd, $2 j \pm s$ is odd, hence $\sin(\pi (2 j \pm s)/2)$  is $-1$ or 1. If $g_{j,s} = 0$, then the brackets must be zero, which would imply 
\begin{align*}
     \frac{\sin(\pi (2 j - s)/2)}{2 j - s} = \frac{\sin(\pi (2 j + s)/2)}{2 j + s}.
\end{align*}
Since both numerators are $\pm 1$, this forces $|2 j - s| = |2 j + s|$, which is not possible for $s \geq 1$. Therefore, $g_{j,s} > 0$ for all $j$ when $s$ is odd.
Finally, 
\begin{align*}
    \left| \frac{\sin(\pi (2 j - s)/2)}{2 j - s} - \frac{\sin(\pi (2 j + s)/2)}{2 j + s} \right| \leq \frac{1}{|2 j - s|} + \frac{1}{|2 j + s|}  = O(j^{-1})
\end{align*}
so $g_{j,s} = O(j^{-2})$ as $j \rightarrow \infty$.

\noindent\textbf{Case $s$ is even.} Assume $s$ is even. Then $\sin(\pi (2 j \pm s)/2) = 0$ for all $j$. 

If $j \neq s/2$ then $\sin(\pi (2 j \pm s)/(2T)) \neq 0$ so 
\begin{align*}
    I_{P(s)}(\lambda_j) = 0.
\end{align*}

If $j = s/2$, then $\sin(\pi (2 j - s)/(2T)) = 0$. Hence, the ratio corresponding to $2j-s = 0$ in $I_{P(s)}(\lambda_{s/2})$ is $0/0$ and therefore must be interpreted as a limit. From the formula of $ w_{P(s)}(\lambda_j)$ in \eqref{w_P} note that at $j = s/2$ the sum with $s \pi/T + \lambda_j$ equals zero, while the sum with $\lambda_j-s \pi/T $ equals $T$. Hence
\begin{align*}
     w_{P(s)}(\lambda_{s/2}) = \frac{1}{2\sqrt{\pi T}} e^{-i\frac{s\pi}{2T}} T
\end{align*}
thus
\begin{align*}
     I_{P(s)}(\lambda_{s/2}) = T \frac{1}{4\pi}.
\end{align*}
Therefore, the representation $I_{P(s)}(\lambda_j) = T g_{j,s} (1+o(1))$ holds with $g_{j,s} = 0$   for $j \neq s/2$ and  $g_{s/2,s} = 1/(4\pi) > 0$. In particular, $g_{j,s} = O(j^{-2})$ for $j \rightarrow \infty$. 
\end{proof}

\begin{lemma}\label{lemma:wtimesu}
    Let $y_t$ be generated according to \eqref{eq1}--\eqref{eq21}, and let Assumption \ref{ass2} hold. Let $m \rightarrow \infty$ and $m = o(T)$. Then for any integer $s \geq 1$ 
    \begin{align*}
    m^{-1} \sum_{j = 1}^{m}  \lambda_j^{2 \delta_0} Re(w_{u}(\lambda_j) w_{P(s)}(-\lambda_j))   = O_p( m^{-1} T^{1/2-\delta_0}).
\end{align*}
\end{lemma}
\begin{proof}
It is sufficient to show that
\begin{align*}
 \Var\left(m^{-1} \sum_{j = 1}^{m}  \lambda_j^{2\delta_0} w_{u}(\lambda_j) w_{P(s)}(-\lambda_j)\right) = O_p( m^{-2} T^{1-2\delta_0}).   
\end{align*}
We introduce the notation as in \textcite{shao2007local}. Let  $g_j = w_{u}(\lambda_j)/(|\alpha(\lambda_j)|\sqrt{f_{\eta}(\lambda_j)}) $
and $h_j = w_{\eta}(\lambda_j)/\sqrt{f_{\eta}(\lambda_j)}$ where $\alpha(\lambda_j) =  (1-e^{i \lambda_j})^{-\delta_0}$.
Set $z_j = g_j - \frac{\alpha(-\lambda_j)}{|\alpha(\lambda_j)|} h_j$. Then
\begin{align*}
    w_{u} = \alpha(\lambda_j)w_{\eta}(\lambda_j) + |\alpha(\lambda_j)| \sqrt{f_{\eta}(\lambda_j)} z_j
\end{align*}
Lemma 3 of \textcite{shao2007local} implies 
\begin{align}
E\left|z_j \right|^2 = O(\ln(j)/j).   \label{usefullbound}
\end{align}
Substituting this expression for $w_{u}(\lambda_j)$ yields
\begin{align}
    m^{-1} \sum_{j = 1}^{m}  \lambda_j^{2\delta_0}  w_{P(s)}(-\lambda_j) w_{u}(\lambda_j) &=  m^{-1} \sum_{j = 1}^{m}  \lambda_j^{2\delta_0}  w_{P(s)}(-\lambda_j)  \alpha(\lambda_j)w_{\eta}(\lambda_j) \nonumber \\
    &\ \ \ +  m^{-1} \sum_{j = 1}^{m}  \lambda_j^{2\delta_0}  w_{P(s)}(-\lambda_j)  |\alpha(\lambda_j)| \sqrt{f_{\eta}(\lambda_j)} z_j \nonumber\\
    &= m^{-1}  \sum_{j = 1}^{m} b_{1,j} w_{\eta}(\lambda_j) +  m^{-1}  \sum_{j = 1}^{m} b_{2,j} z_j, \label{sumtb}
\end{align}
where $b_{1,j} = \lambda_j^{2\delta_0}  w_{P(s)}(-\lambda_j)  \alpha(\lambda_j) $ and $b_{2,j} =  \lambda_j^{2\delta_0}  w_{P(s)}(-\lambda_j)  |\alpha(\lambda_j)| \sqrt{f_{\eta}(\lambda_j)}$. 

Next, note that $|1-e^{i\lambda_j}| \sim  C\lambda_j$. Hence $\lambda^{\delta_0}_j|\alpha(\lambda_j)|$ is bounded, so $|\alpha(\lambda_j)| \leq C \lambda^{-\delta_0}_j$. By Lemma \ref{w_P}, $I_{P(s)}(\lambda_j) \leq C T j^{-2}$. It follows that $|b_{1,j}| \leq C T^{1/2-\delta_0} j^{-1+\delta_0}$. Since  $f_{\eta}(\lambda_j)$ is bounded under Assumption \ref{ass2}, the same bound hold for $b_{2,j}$, i.e.\ $|b_{2,j}| \leq C T^{1/2-\delta_0} j^{-1+\delta_0}$.

We first bound the variance of the first summand in \eqref{sumtb}.
From Fact 3 in \textcite{shao2007local} $\Cov( w_{\eta}(\lambda_j),  w_{\eta}(-\lambda_k)) = f_{\eta}(\lambda_j) 1(j = k) + O(T^{-1})$, where $f_{\eta}(\lambda_j) < C$.
Therefore, 
\begin{align*}
   \Var\left(  m^{-1}  \sum_{j = 1}^{m} b_{1,j} w_{\eta}(\lambda_j) \right)  &= m^{-2}  \sum_{j = 1}^{m}  \sum_{k = 1}^{m}  b_{1,j} b_{1,k} \Cov( w_{\eta}(\lambda_j),  w_{\eta}(-\lambda_k)) \\
   &\leq C m^{-2}  \sum_{j = 1}^{m} |b_{1,j}|^2 + C T^{-1}m^{-2} \left(  \sum_{j = 1}^{m} |b_{1,j}| \right)^2.
\end{align*}
Using the bound $|b_{1,j}| \leq C T^{1/2-\delta_0} j^{-1+\delta_0}$ and $|\delta_0| \leq 1/2$,
\begin{align*}
    m^{-2} \sum_{j = 1}^m |b_{1,j}|^2 \leq  C  m^{-2} T^{1-2\delta_0} \sum_{j = 1}^{m}  j^{-2+2\delta_0} =   O( m^{-2} T^{1-2\delta_0})
\end{align*}
and by Cauchy-Schwarz,
\begin{align*}
    \left(  \sum_{j = 1}^{m} |b_{1,j}| \right)^2 \leq m \sum_{j = 1}^m |b_{1,j}|^2 = O( m T^{1-2\delta_0}).
\end{align*}
Hence,
\begin{align*}
     \Var\left(  m^{-1}  \sum_{j = 1}^{m} b_{1,j} w_{\eta}(\lambda_j) \right)  &\leq C  m^{-2} T^{1-2\delta_0} +  C m^{-2} T^{1-2\delta_0} m T^{-1}\\
     &\leq O( m^{-2} T^{1-2\delta_0}) + O(m^{-2} T^{1-2\delta_0} T^{-1} m) = O( m^{-2} T^{1-2\delta_0}). 
\end{align*}
Since $m = o(T)$, the second term is $o(m^{-2}T^{1-2\delta_0})$, so 
\begin{align*}
    \Var\left(  m^{-1}  \sum_{j = 1}^{m} b_{1,j} w_{\eta}(\lambda_j) \right) = O( m^{-2} T^{1-2\delta_0}).
\end{align*}

Next, we bound the variance of the second summand in \eqref{sumtb}.
By Cauchy-Schwarz, $|b_{j,2}| \leq C T^{1/2-\delta_0} j^{-1+\delta_0}$, and \eqref{usefullbound},
\begin{align*}
     \Var\left(m^{-1}  \sum_{j = 1}^{m} b_{2,j} z_j\right) &\leq E\left|m^{-1}  \sum_{j = 1}^{m} b_{2,j} z_j \right|^2 \\
     &\leq m^{-2} \left( \sum_{j = 1}^{m} |b_{2,j}| \sqrt{E |z_j|^2} \right)^2 \\
     &\leq  C T^{1-2\delta_0} m^{-2} \left( \sum_{j = 1}^{m} j^{-3/2+\delta_0} (\ln(j))^{1/2} \right)^2.
\end{align*}
Because $|\delta_0| < 1/2$, we have $\sum_{j = 1}^{m} j^{-3/2+\delta_0} (\ln(j))^{1/2} = O(1)$, and therefore
\begin{align*}
    \Var\left(m^{-1}  \sum_{j = 1}^{m} b_{2,j} z_j\right) = O(T^{1-2\delta_0} m^{-2}).
\end{align*}

Finally, since $Z = Z_1 + Z_2$ we use $\Var(Z_1 + Z_2) \leq 2 \Var(Z_1) + 2\Var(Z_2)$ to obtain 
\begin{align*}
     \Var\left(m^{-1} \sum_{j = 1}^{m}  \lambda_j^{2\delta_0} w_{u}(\lambda_j) w_{P(s)}(-\lambda_j)\right) &\leq 2 \Var\left(  m^{-1}  \sum_{j = 1}^{m} b_{1,j} w_{\eta}(\lambda_j) \right) + 2 \Var\left(  m^{-1}  \sum_{j = 1}^{m} b_{2,j} z_j \right) \\
    &= O(T^{1-2\delta_0} m^{-2}).
\end{align*}
Since 
\begin{align*}
    \Var \left( m^{-1} \sum_{j = 1}^{m}  \lambda_j^{2 \delta_0} Re(w_{u}(\lambda_j) w_{P(s)}(-\lambda_j)) \right) \leq  \Var\left(m^{-1} \sum_{j = 1}^{m}  \lambda_j^{2\delta_0} w_{u}(\lambda_j) w_{P(s)}(-\lambda_j)\right),
\end{align*}
the desired result follows.
\end{proof}

\begin{lemma} \label{thm6}
 Let $y_t$ be generated according to \eqref{eq1}--\eqref{eq21} and let Assumptions \ref{ass2} and \eqref{ass6} hold. Then for any fixed $k \geq k_0$,
\begin{align*}
\hat\delta(k) \xrightarrow{p}\delta_0, 
\end{align*}
and
\begin{align*}
    \sqrt{m}\big(\hat\delta(k)-\delta_0\big) \xrightarrow{d} \mathcal N(0,1/4).
\end{align*}
\end{lemma}
\begin{proof}
Fix $k \geq k_0$. Then $\hat{u}_t(k) = u_t - \sum_{i = 0}^{k} (\hat{\beta}_i - \beta_{i,0})P_t(i)$ and, for $j = 1,\ldots,m$, 
\begin{align*}
    I_{\hat{u}(k)}(\lambda_j) &=  I_{u}(\lambda_j) + \sum_{i = 1}^k \left(\hat{\beta}_i - \beta_{i,0} \right)^2 I_{P(i)}(\lambda_j)-2 \sum_{i = 1}^k \left(\hat{\beta}_i - \beta_{i,0} \right) Re(w_{u}(\lambda_j) w_{P(i)}(-\lambda_j) ) \\
    & \ \ \ + 2 \sum_{s = 1}^{k-1} \sum_{i = s+1}^{k} \left(\hat{\beta}_s - \beta_{s,0} \right)\left(\hat{\beta}_i - \beta_{i,0} \right)  Re(w_{P(s)}(\lambda_j) w_{P(i)}(-\lambda_j)).
\end{align*}
Let $\eta_j = \lambda_j^{2 \delta_0} I_{\hat{u}(k)}(\lambda_j)/G_0$ and $\eta^0_j = \lambda_j^{2 \delta_0} I_{u}(\lambda_j)/G_0$. By Proposition 1 of \textcite{dalla2006consistent}, $\hat{\delta}(k) \overset{p}{\rightarrow} d_0$, provided that the following two requirements hold:
\begin{align*}
    E(\eta_j) \leq C,
\end{align*}
for $j = 1,\ldots,m$ and that
\begin{align}
    \frac{1}{\lfloor \tau m\rfloor} \sum_{j = 1}^{\lfloor \tau m\rfloor} \eta_j \overset{p}{\rightarrow} 1,  \label{peta}
\end{align}
for any $\tau \in (0,1]$.

We first consider the first requirement.
By Proposition 3.1 of \textcite{shao2007local1},
\begin{align*}
    E\left( \eta_j^0 \right) = \frac{\lambda_j^{2\delta_0}}{G_0} E I_u(\lambda_j) \leq C,
\end{align*}
uniformly for $1\leq j \leq m$. For the remaining terms, Lemma \ref{consCheby} implies $E(\hat{\beta}_i - \beta_{i,0})^2 = O(T^{-1+2\delta_0})$ for each $i \leq k$, Lemma \ref{lemma:w_P} yields $I_{P(i)}(\lambda_j) \leq C T^{-2}$. 

First, 
\begin{align*}
    \lambda_j^{2 \delta_0} I_{P(i)} (\lambda_j)  E(\hat{\beta}_i - \beta_{i,0})^2  &\leq C (j/T)^{2\delta_0} T j^{-2} T^{-1+2\delta_0} \leq C j^{2\delta_0-2} \leq C.
\end{align*}
Next, by Cauchy-Schwarz and $|Re(z)| \leq z$,
\begin{align*}
    \lambda_j^{2 \delta_0} \left| E \left[(\hat{\beta}_i - \beta_{i,0})  Re(w_{u}(\lambda_j) w_{P(i)}(-\lambda_j)) \right] \right| &\leq   \lambda_j^{2 \delta_0} \sqrt{E(\hat{\beta}_i - \beta_{i,0})^2} \sqrt{ I_{P(i)}(\lambda_j) }  \sqrt{E I_{u}(\lambda_j)} \\
    &\leq C j^{\delta_0-1} \leq C.
\end{align*}
Finally, by Cauchy-Schwarz and $|Re(z)| \leq z$,
\begin{align*}
    \lambda_j^{2 \delta_0} \left|Re(w_{P(s)}(\lambda_j) w_{P(i)}(-\lambda_j))  E\left[(\hat{\beta}_s - \beta_{s,0}) (\hat{\beta}_i - \beta_{i,0}) \right] \right| &\leq \lambda_j^{2 \delta_0} \sqrt{I_{P(s)}(\lambda_j)} \sqrt{I_{P(i)}(\lambda_j)}  \\ 
    & \ \ \ \times \sqrt{E(\hat{\beta}_s - \beta_{s,0})^2} \sqrt{E(\hat{\beta}_i - \beta_{i,0})^2} \\
    &\leq C j^{2\delta_0-2} \leq C.
\end{align*}
Combining these bounds yields $E(\eta_j) \leq C$ uniformly for $j = 1,\ldots,m$. 

For the second requirement, write 
\begin{align*}
    \frac{1}{\lfloor \tau m\rfloor} \sum_{j = 1}^{\lfloor \tau m\rfloor} \eta_j =  \frac{1}{\lfloor \tau m\rfloor} \sum_{j = 1}^{\lfloor \tau m\rfloor} \eta^0_j + \frac{1}{\lfloor \tau m\rfloor} \sum_{j = 1}^{\lfloor \tau m\rfloor} \frac{\lambda_j^{2\delta_0}}{G_0}  \left(I_{\hat{u}(k)}(\lambda_j) - I_{u}(\lambda_j) \right).
\end{align*}
In the proof of Proposition 3.1 in \textcite{shao2007local1}, it is shown that
\begin{align}
    \frac{1}{\lfloor \tau m\rfloor} \sum_{j = 1}^{\lfloor \tau m\rfloor} \eta^0_j  \overset{p}{\rightarrow} 1. \label{peta0}
\end{align}
Moreover, the above bounds imply that each term in $\lambda_j^{2\delta_0}  (I_{\hat{u}(k)}(\lambda_j) - I_{u}(\lambda_j))$ is dominated by a constant times the summation of $j^{2\delta_0-2}$, expect for the term involving $Re(w_{u}(\lambda_j) w_{P(i)}(-\lambda_j) )$. By Lemma \ref{lemma:wtimesu}, this term is $O_p((\lfloor \tau m\rfloor)^{-1})$. Hence 
\begin{align}
    E \left| \frac{1}{\lfloor \tau m\rfloor} \sum_{j = 1}^{\lfloor \tau m\rfloor} \frac{\lambda_j^{2\delta_0}}{G_0}  \left(I_{\hat{u}(k)}(\lambda_j) - I_{u}(\lambda_j) \right)\right| \leq C \frac{1}{\lfloor \tau m\rfloor} \sum_{j = 1}^{\lfloor \tau m\rfloor} j^{2\delta_0 - 2} + C \frac{1}{\lfloor \tau m\rfloor}   = O\left( (\lfloor \tau m\rfloor)^{-1}  \right) = o(1), \label{boundRd_0}
\end{align}
and therefore $(\lfloor \tau m\rfloor)^{-1} \sum_{j = 1}^{\lfloor \tau m\rfloor} \eta_j \overset{p}{\rightarrow} 1$.

Now we show $\sqrt{m}\big(\hat\delta(k)-\delta_0\big) \xrightarrow{d} \mathcal N(0,1/4)$. By Proposition 1 of \textcite{dalla2006consistent}, since the two requirements hold, 
\begin{align}
    2 \left(\hat{\delta}(k) - \delta_0 \right) = - F_m (1+o_p(1)) + O_p(\ln(m)/m), \label{expansiond}
\end{align}
where $F_m = m^{-1} \sum_{j = 1}^m s_j \lambda_j^{2\delta_0} G^{-1}_0 I_{\hat{u}(k)}(\lambda_j)$ and $s_j = 1 + \ln(j/m)$. Hence it suffices to show 
\begin{align*}
    \sqrt{m} F_m \xrightarrow{d} \mathcal N(0,1).
\end{align*}
Let $F^0_m = m^{-1} \sum_{j = 1}^m s_j \lambda_j^{2\delta_0} G^{-1}_0 I_{u}(\lambda_j)$. By \textcite[Theorem 3.1]{shao2007local1}, 
\begin{align*}
    \sqrt{m} F^0_m \xrightarrow{d} \mathcal N(0,1).
\end{align*}
Moreover, using the decomposition of $I_{\hat{u}(k)}(\lambda_j)$ and the bounds used earlier, it can be shown that $m^{1/2}(F_m-F_m^0) = o_p(1)$, so $\sqrt{m} F_m \xrightarrow{d} \mathcal N(0,1)$.

\end{proof}

\subsection{Proof of Theorem \ref{thm1}}
\begin{proof}
For simplicity, we provide the proof for the case where $k = 1$ and $k_0 = 2$. The proof of the general case follows by analogous arguments, with the only inconvenience being the introduction of additional notation. 

Recall that for our second claim we assume that $T^{1-2\delta_0}  \ln(m) m^{-1/2} \rightarrow \infty$ and $T^{1-2\delta_0}   m^{-1} \rightarrow \infty$. It then suffices to show that 
\begin{align}
    m^{-1/2}\sum_{j = 1}^m v_j \lambda_j^{2 \delta_0} I_{\hat{u}(k)}(\lambda_j) \overset{p}{\sim} -C_1 T^{1-2\delta_0}  \ln(m) m^{-1/2}, \label{s1} \\ 
   m^{-1}\sum_{j = 1}^m \lambda_j^{2 \delta_0} I_{\hat{u}(k)}(\lambda_j) \overset{p}{\sim} C_2  T^{1-2\delta_0}   m^{-1}, \label{s22}
\end{align} 
where $C_1$ and $C_2$ are positive constants. For the first claim, we only require $T^{1-2\delta_0}  \ln(m) m^{-1/2} \rightarrow \infty$. Since the denominator in \eqref{s22} is strictly positive (as a sum of weighted periodograms), and
\begin{align*}
    \frac{T^{1-2\delta_0}m^{-1}}{T^{1-2\delta_0}\ln(m)m^{-1/2}}
=\frac{1}{\sqrt{m}\ln(m)} \to 0,
\end{align*}
divergence of $t_{\hat u(k)}(\delta_0;m)$ and $\LM_{\hat u(k)}(\delta_0;m)$ under $H_0$ follows once \eqref{s1} is shown under the bandwidth condition $T^{1-2\delta_0}\ln(m)m^{-1/2}\rightarrow\infty$.

We provide the proof of \eqref{s1} only. The proof of \eqref{s22} follows from similar arguments, noting the absence of $v_j$ and the different normalisation. 

Note that the fitted residual is 
\begin{align*}
    \hat{u}_t(k) &= \beta_{0,0} P_{t}(0) + \beta_{1,0} P_{t}(1) + \beta_2 P_{t}(2) + u_t - \hat{\beta}_0 P_{t}(0) - \hat{\beta}_1 P_{t}(1)  \\
    &= u_t + (\beta_{0,0} - \hat{\beta}_0) P_{t}(0) + (\beta_{1,0} - \hat{\beta}_1)  P_{t}(1)  + \beta_{2,0}P_{t}(2). 
\end{align*}
Since $P_t(0) = \sqrt{2}$ is constant in $t$, its discrete Fourier transform satisfies $w_{P(0)}(\lambda_j) = 0$ for $j \geq 1$. Therefore, all periodogram terms involving $P_t(0)$ vanish and we may ignore the $(\beta_{0,0} - \hat{\beta}_0) P_{t}(0)$ components.

Using $I_x(\lambda_j) = |w_x(\lambda_j)|^2$ and $|a+b|^2 = |a|^2 + |b|^2 + 2 Re(a \bar{b})$, we obtain for $j \geq 1$:
\begin{align}
    I_{\hat{u}(k)}(\lambda_j) &=  I_{u}(\lambda_j) + (\beta_{1,0} - \hat{\beta}_1)^2  I_{P(1)}(\lambda_j) + \beta^2_{2,0}   I_{P(2)}(\lambda_j) \nonumber \\
    &\quad + 2 (\beta_{1,0} - \hat{\beta}_1) Re(w_{u}(\lambda_j) w_{P(1)}(-\lambda_j)   ) \nonumber \\
    &\quad +  2 \beta_{2,0} Re(w_{u}(\lambda_j) w_{P(2)}(-\lambda_j)   ) \nonumber \\
    &\quad + 2 (\beta_{1,0} - \hat{\beta}_1) \beta_{2,0} Re(w_{P(1)}(\lambda_j) w_{P(2)}(-\lambda_j)   ). \label{Idec}
\end{align}
Hence, \eqref{s1} follows if the following bounds hold: 
\begin{align}
    m^{-1/2}\sum_{j = 1}^m v_j \lambda_j^{2 \delta_0} I_{u}(\lambda_j) &= O_p(1), \label{a1} \\
    (\beta_{1,0} - \hat{\beta}_1)^2  m^{-1/2}\sum_{j = 1}^m v_j \lambda_j^{2 \delta_0}  I_{P(1)}(\lambda_j) &= o_p(1), \label{a2} \\
    \beta^2_{2,0}   m^{-1/2}\sum_{j = 1}^m v_j \lambda_j^{2 \delta_0} I_{P(2)}(\lambda_j)  &\sim  -C_1 T^{1-2\delta_0}   \ln(m) m^{- 1/2} ,  \label{a3} \\
    2 (\beta_{1,0} - \hat{\beta}_1) m^{-1/2}\sum_{j = 1}^m v_j \lambda_j^{2 \delta_0} Re(w_{u}(\lambda_j) w_{P(1)}(-\lambda_j)) &= o_p(1), \label{a4} \\
    2 \beta_{2,0} m^{-1/2}\sum_{j = 1}^m v_j \lambda_j^{2 \delta_0} Re(w_{u}(\lambda_j) w_{P(2)}(-\lambda_j)   )  &= O_p\left( T^{1/2-\delta_0} \ln(m) m^{-1/2}   \right),  \label{a5} \\
      2 (\beta_{1,0} - \hat{\beta}_1) \beta_{2,0} m^{-1/2}\sum_{j = 1}^m v_j \lambda_j^{2 \delta_0}  Re(w_{P(1)}(\lambda_j) w_{P(2)}(-\lambda_j)) &= O_p\left( T^{1/2-\delta_0}\ln(m) m^{ - 1/2}\right).  \label{a6} 
\end{align}
Note that the leading term in \eqref{a3} is of order $T^{1-2\delta_0} \ln(m) m^{-1/2} \rightarrow \infty$, which dominates the orders in \eqref{a1}, \eqref{a2}, \eqref{a4} and also dominates \eqref{a5}-\eqref{a6} by a factor $T^{-1/2+\delta_0}$.  

\textit{Proof of \eqref{a1}:} By Theorem 1 in \textcite{iacone2022semiparametric} and the arguments in their proof, the quantity $ m^{-1/2}\sum_{j = 1}^m v_j \lambda_j^{2 \delta_0} I_{u}(\lambda_j)$, normalised by $G$, converges in distribution to a standard normal. Hence it is $O_p(1)$.

\textit{Proof of \eqref{a2}:} 
By Lemma \ref{consCheby}, $(\beta_{1,0} - \hat{\beta}_1)^2 = O_p(T^{2\delta_0 - 1})$.
By Lemma \ref{lemma:w_P}, for any fixed $s\geq 1$,uniformly for $1 \leq j \leq m$: 
\begin{align}
    I_{P(s)}(\lambda_j) \leq C T j^{-2}, \label{boundp}
\end{align}
Moreover, $|v_j| = O(\ln(m))$ uniformly for $1 \leq j \leq m$.
Therefore, 
\begin{align*}
   (\beta_{1,0} - \hat{\beta}_1)^2  m^{-1/2}\sum_{j = 1}^m v_j \lambda_j^{2 \delta_0}  I_{P(1)}(\lambda_j) &= O_p( T^{2\delta_0 - 1} m^{-1/2} \sum_{j = 1}^m \ln(m) (j/T)^{2 \delta_0}   T j^{-2} ) \\
      &= O_p(m^{-1/2}  \ln(m) \sum_{j = 1}^m j^{2 \delta_0 -2}) \\
      &= O_p(m^{-1/2}  \ln(m)  )   = o_p(1).
\end{align*}

\textit{Proof of \eqref{a3}:} By Lemma \ref{lemma:w_P}, for $s$ = 2 (even) we have
\begin{align*}
 I_{P(2)}(\lambda_j) = Tg_{j,2}(1+o(1)),
\end{align*}
uniformly for $1 \leq j \leq m$, with $g_{j,2} = 0$ for $j \neq 1$ and $g_{1,2} > 0$. Hence the sum in \eqref{a3} collapses to the single term $j = 1$:  
\begin{align*}
     \beta^2_{2,0}   m^{-1/2}\sum_{j = 1}^m v_j \lambda_j^{2 \delta_0} I_{P(2)}(\lambda_j) =  \beta^2_{2,0}   m^{-1/2} v_1  \lambda_j^{2 \delta_0}   Tg_{j,2}(1+o(1)).
\end{align*}
Since $\lambda_1 = 2\pi/T$, this equals
\begin{align*}
\beta^2_{2,0} g_{1,2} (2\pi)^{2\delta_0} T^{1-2\delta_0}   m^{-1/2} v_1(1+o(1)).
\end{align*}
By definition 
\begin{align*}
    v_1 = \ln(1)- \frac{1}{m} \sum_{j = 1}^m \ln(j) = -\frac{1}{m} \sum_{j = 1}^m \ln(j) =  -\frac{1}{m} \ln(m!)
\end{align*}
Using Stirling's approximation $\ln(m!) = m \ln(m) - m + O(\ln(m))$, we obtain
\begin{align*}
    v_1 = -\ln(m)(1+o(1)).
\end{align*}
Substituting this back yields
\begin{align*}
        \beta^2_{2,0}   m^{-1/2}\sum_{j = 1}^m v_j \lambda_j^{2 \delta_0} I_{P(2)}(\lambda_j) \sim -C_1 T^{1-2\delta_0} \ln(m)   m^{-1/2},
\end{align*}
for $C_1 = \beta^2_{2,0} g_{1,2} (2\pi)^{2\delta_0} >0$.

\textit{Proof of \eqref{a4}:}
Note
\begin{align*}
    |Re(w_{u}(\lambda_j) w_{P(1)}(-\lambda_j))| \leq |w_{u}(\lambda_j)| | w_{P(1)}(-\lambda_j)|.
\end{align*}
From \eqref{boundp}, $|w_{P(1)(\lambda_j)}| \leq C T^{1/2} j^{-1}$ uniformly for $1\leq j \leq m$. Moreover, $E\left| \lambda_j^{2\delta_0} I_u(\lambda_j)  \right| <C$ by Lemma 3 in \textcite{shao2007local}, so $|w_u(\lambda_j)| = O_p(\lambda_j^{-\delta_0})$ uniformly for  $1\leq j \leq m$.

Therefore, uniformly for  $1\leq j \leq m$,
\begin{align}
     \lambda_j^{2\delta_0} |Re(w_{u}(\lambda_j) w_{P(1)}(-\lambda_j))| &\leq  \lambda_j^{2\delta_0}  |w_{u}(\lambda_j)| | w_{P(1)}(-\lambda_j)| \nonumber\\ &= O_p(\lambda_j^{\delta_0} T^{1/2} j^{-1}) = O_p(T^{1/2-\delta_0} j^{-1+\delta_0}). \label{bound1}
\end{align}

By Lemma \ref{consCheby}, $( \hat{\beta}_1-\beta_{1,0}) = O_p(T^{\delta_0-1/2})$ and using $|v_j| = O(\ln(m))$  
\begin{align*}
      (\beta_{1,0} - \hat{\beta}_1) m^{-1/2}\sum_{j = 1}^m v_j \lambda_j^{2 \delta_0} Re(w_{u}(\lambda_j) w_{P(1)}(-\lambda_j)) &= O_p(  T^{\delta_0-1/2} m^{-1/2}\sum_{j = 1}^m  \ln(m)  T^{1/2-\delta_0} j^{-1+\delta_0}  ) \\
      &= O_p(m^{-1/2} \ln(m) \sum_{j = 1}^m j^{-1+\delta_0} ) \\
      &= O_p(m^{-1/2+\max(0,\delta_0)} \ln^2(m)) = o_p(1),
\end{align*}
since $|\delta_0| < 0.5$. 

\textit{Proof of \eqref{a5}:} By Lemma \ref{lemma:w_P}, $w_{P(2)}(\lambda_j) = 0$ for $j \neq 1$, so the sum reduces to $j = 1$:
\begin{align*}
    \beta_{2,0} m^{-1/2}\sum_{j = 1}^m v_j \lambda_j^{2 \delta_0} Re(w_{u}(\lambda_j) w_{P(2)}(-\lambda_j)) = \beta_{2,0} m^{-1/2} v_1 \lambda_1^{2 \delta_0} Re(w_{u}(\lambda_1) w_{P(2)}(-\lambda_1)).
\end{align*}
Using the bound in \eqref{bound1} and $|v_j| = O(\ln(m))$ yields
\begin{align*}
   \beta_{2,0} m^{-1/2}\sum_{j = 1}^m v_j \lambda_j^{2 \delta_0} Re(w_{u}(\lambda_j) w_{P(2)}(-\lambda_j)) = O_p(m^{-1/2} \ln(m) T^{1/2-\delta_0}).
\end{align*}

\textit{Proof of \eqref{a6}:} 
By Lemma \ref{consCheby}, $( \hat{\beta}_1-\beta_{1,0}) = O_p(T^{\delta_0-1/2})$. Using the bound in \eqref{boundp} and $|v_j| = O(\ln(m))$ yields
\begin{align*}
    (\beta_{1,0} - \hat{\beta}_1) \beta_{2,0} m^{-1/2}\sum_{j = 1}^m v_j \lambda_j^{2 \delta_0}  Re(w_{P(1)}(\lambda_j) w_{P(2)}(-\lambda_j)) &= O_p(  T^{\delta_0-1/2}  m^{-1/2}\sum_{j = 1}^m \ln(m) (j/T)^{2 \delta_0}  T j^{-2}     ) \\
      &= O_p(  T^{1/2-\delta_0} m^{-1/2} \ln(m) \sum_{j = 1}^m j^{-2+2\delta_0} ) \\ 
      &= O_p( T^{1/2-\delta_0} m^{-1/2} \ln(m)),
\end{align*}
since $|\delta_0| < 0.5$.

\end{proof}

\subsection{Proof of Theorem \ref{thm2}}

\begin{proof}
Following a similar proof strategy as in  \textcite{iacone2022semiparametric}, we divide the proof into two parts. First, we consider the case $k = k_0$. Afterwards, we consider the case $k > k_0$. When the true errors $u_t$ are observed, \textcite[Theorem 1]{iacone2022semiparametric} show that, under the local alternative $\delta = \delta_0 + c m^{-1/2}$, $t_{u}(\delta_0;m)   \overset{d}{\rightarrow} N(2c,1)$. Using this result, it remains to show that $ t_{\hat{u}(k)}(\delta_0;m) - t_{u}(\delta_0;m) \overset{p}{\rightarrow} 0$  under the same local alternative $\delta = \delta_0 + c m^{-1/2}$. 

\textit{When $k = k_0$}. We only consider $k = k_0 = 1$ for simplicity. In this case, the true coefficient vector has no omitted Chebyshev terms beyond order 1. 

It is sufficient to show that 
\begin{align}
     m^{-1/2}\sum_{j = 1}^m v_j \lambda_j^{2 \delta_0} I_{\hat{u}(k)}(\lambda_j) -  m^{-1/2}\sum_{j = 1}^m v_j 
    \lambda_j^{2 \delta_0} I_{u}(\lambda_j) = o_p(1), \label{b1} \\
       m^{-1}\sum_{j = 1}^m \lambda_j^{2 \delta_0} I_{\hat{u}(k)}(\lambda_j) -  m^{-1}\sum_{j = 1}^m 
    \lambda_j^{2 \delta_0} I_{u}(\lambda_j) = o_p(1). \label{b2} 
\end{align}
We only provide the proof of \eqref{b1}. The proof of \eqref{b2} follows from identical arguments (without the weights $v_j$ and with a different normalisation). 

We have 
\begin{align*}
    \hat{u}_t(k) &= \beta_{0,0} P_{t}(0)  + \beta_{1,0} P_{t}(1) + u_t - \hat{\beta}_0 P_{t}(0) - \hat{\beta}_1 P_{t}(1)  \\
    &= u_t + (\beta_{0,0} - \hat{\beta}_0) P_{t}(0) + (\beta_{1,0} - \hat{\beta}_1)  P_{t}(1), 
\end{align*}
and hence, for $j \geq 1$, 
\begin{align*}
    I_{\hat{u}(k)}(\lambda_j) &=  I_{u}(\lambda_j) + (\beta_{1,0} - \hat{\beta}_1)^2  I_{P(1)}(\lambda_j) + 2 (\beta_{1,0} - \hat{\beta}_1) Re(w_{u}(\lambda_j) w_{P(1)}(-\lambda_j) ). 
\end{align*}
Therefore, the absolute value of the left-hand side of \eqref{b1} is bounded by
\begin{align*}
    (\beta_{1,0} - \hat{\beta}_1)^2  m^{-1/2}\sum_{j = 1}^m |v_j| \lambda_j^{2 \delta_0}  I_{P(1)}(\lambda_j) + 2 |(\beta_{1,0} - \hat{\beta}_1)|  m^{-1/2}\sum_{j = 1}^m |v_j| \lambda_j^{2 \delta_0} |Re(w_{u}(\lambda_j) w_{P(1)}(-\lambda_j))|. 
\end{align*}
Thus, \eqref{b1} holds if the following two conditions are satisfied: 
\begin{align}
    (\beta_{1,0} - \hat{\beta}_1)^2  m^{-1/2}\sum_{j = 1}^m |v_j| \lambda_j^{2 \delta_0}  I_{P(1)}(\lambda_j) = o_p(1),\label{b3} \\
    2 |(\beta_{1,0} - \hat{\beta}_1)|  m^{-1/2}\sum_{j = 1}^m |v_j| \lambda_j^{2 \delta_0} |Re(w_{u}(\lambda_j) w_{P(1)}(-\lambda_j))| = o_p(1),\label{b4}
\end{align}

First, by Lemma \ref{consCheby}, $(\hat{\beta}_1-\beta_{1,0}) = O_p(T^{\delta_0 - 1/2})$. The proofs of \eqref{b3} then follows exactly as in the proofs in \eqref{a2}, using the bound $I_{P(s)} \leq C T j^{-2}$ and $|v_j| = O(\ln(m))$.

We next prove \eqref{b4}. Under the local alternative $\delta = \delta_0 + c m^{-1/2}$, Lemma 3 in \textcite{shao2007local} show $E\left| \lambda_j^{2\delta} I_u(\lambda_j)  \right| <C$ uniformly for $1 \leq j \leq m $ and hence $|w_u(\lambda_j)| = O_p(\lambda_j^{-\delta})$. Moreover, from the bound in \eqref{bound1}, we have $|w_{P(s)}(\lambda_j)| = O_p(T^{1/2}j^{-1})$.
Therefore, uniformly for  $1\leq j \leq m$ and any $s \geq 1$,
\begin{align}
     \lambda_j^{2\delta_0} |Re(w_{u}(\lambda_j) w_{P(s)}(-\lambda_j))| &\leq  \lambda_j^{2\delta_0}  |w_{u}(\lambda_j)| | w_{P(s)}(-\lambda_j)| \nonumber\\ &= O_p(\lambda_j^{2\delta_0-\delta} T^{1/2} j^{-1}) = O_p(T^{1/2-2\delta_0+\delta} j^{-1+2\delta_0-\delta}) \label{bound2}
\end{align}
under $\delta = \delta_0 + c m^{-1/2}$.

Using $|v_j| = \ln(m)$ and $(\beta_{1,0} - \hat{\beta}_1) = O_p(T^{\delta_0-1/2})$, we obtain 
\begin{align*}
      |(\beta_{1,0} - \hat{\beta}_1)|  m^{-1/2}\sum_{j = 1}^m |v_j| \lambda_j^{2 \delta_0} |Re(w_{u}(\lambda_j) w_{P(1)}(-\lambda_j))| &= O_p(T^{\delta_0 - 1/2} \ln(m) m^{-1/2}\sum_{j = 1}^m T^{1/2-2\delta_0+\delta} j^{-1+2\delta_0-\delta}) \\
      &= O_p(T^{\delta - \delta_0} \ln(m) m^{-1/2} \sum_{j = 1}^m j^{\delta_0 - 1 + \delta_0 - \delta}).
\end{align*}
As in the treatment of bound (A.15) in the proof of Theorem 2 of \textcite{iacone2022semiparametric}, one may apply a mean-value theorem argument to obtain that $T^{\delta - \delta_0} \rightarrow 1$ and $j^{\delta_0 - \delta}  \rightarrow 1$ under $\delta = \delta_0 + c m^{-1/2}$ . Hence the previous display simplifies to  
\begin{align*}
   O_p( \ln(m) m^{-1/2} \sum_{j = 1}^m j^{\delta_0 - 1}) = O_p( \ln^2(m) m^{-1/2} m^{\max(0,\delta_0)}) = o_p(1),
\end{align*}
since $|\delta_0| < 1/2$.

\textit{When $k > k_0$}. We only consider $k = 2$ and $k_0 = 1$ for simplicity, and the general case follows from similar arguments. In this case, the true coefficient of $P_t(2)$ is zero, i.e.\ $\beta_{2,0} = 0$. It is sufficient to show that 
\begin{align}
     m^{-1/2}\sum_{j = 1}^m v_j \lambda_j^{2 \delta_0} I_{\hat{u}(k)}(\lambda_j) -  m^{-1/2}\sum_{j = 1}^m v_j 
    \lambda_j^{2 \delta_0} I_{u}(\lambda_j) = o_p(1), \label{c1} \\
       m^{-1}\sum_{j = 1}^m \lambda_j^{2 \delta_0} I_{\hat{u}(k)}(\lambda_j) -  m^{-1}\sum_{j = 1}^m 
    \lambda_j^{2 \delta_0} I_{u}(\lambda_j) = o_p(1). \label{c2} 
\end{align}
We only give the proof of \eqref{c1}. The proof of \eqref{c2} follows from identical arguments. 

We have that 
\begin{align*}
    \hat{u}_t(k) &= \beta_{0,0} P_{t}(0) + \beta_{1,0} P_{t}(1) + u_t - \hat{\beta}_0 P_{t}(0) - \hat{\beta}_1 P_{t}(1)  - \hat{\beta}_2 P_{t}(2)  \\
    &= u_t + (\beta_{0,0} - \hat{\beta}_0) P_{t}(0) + (\beta_{1,0} - \hat{\beta}_1)  P_{t}(1)  - \hat{\beta}_2 P_{t}(2), 
\end{align*}
and hence, for $j \geq 1$,
\begin{align*}
    I_{\hat{u}(k)}(\lambda_j) =  &I_{u}(\lambda_j) + (\beta_{1,0} - \hat{\beta}_1)^2  I_{P(1)}(\lambda_j) +  \hat{\beta}_2^2   I_{P(2)}(\lambda_j) \\
    &+ 2 (\beta_{1,0} - \hat{\beta}_1) Re(w_{u}(\lambda_j) w_{P(1)}(-\lambda_j)   ) \\
    &-  2\hat{\beta}_2 Re(w_{u}(\lambda_j) w_{P(2)}(-\lambda_j)   )  \\
    &- 2 (\beta_{1,0} - \hat{\beta}_1) \hat{\beta}_2 Re(w_{P(1)}(\lambda_j) w_{P(2)}(-\lambda_j)   ).
\end{align*}
Therefore, the absolute value of the left-hand side of \eqref{c1} is bounded by 
\begin{align*}
 &(\beta_{1,0} - \hat{\beta}_1)^2  m^{-1/2}\sum_{j = 1}^m |v_j| \lambda_j^{2 \delta_0}  I_{P(1)}(\lambda_j) + \hat{\beta}_2^2   m^{-1/2}\sum_{j = 1}^m |v_j| \lambda_j^{2 \delta_0}  I_{P(2)}(\lambda_j) \\
    &+ 2 |(\beta_{1,0} - \hat{\beta}_1)|  m^{-1/2}\sum_{j = 1}^m |v_j| \lambda_j^{2 \delta_0} |Re(w_{u}(\lambda_j) w_{P(1)}(-\lambda_j))|  \\
    &+    2 |\hat{\beta}_2|  m^{-1/2}\sum_{j = 1}^m |v_j| \lambda_j^{2 \delta_0} |Re(w_{u}(\lambda_j) w_{P(2)}(-\lambda_j))| \\
     &+    2 |(\beta_{1,0} - \hat{\beta}_1)| |\hat{\beta}_2|    m^{-1/2}\sum_{j = 1}^m |v_j| \lambda_j^{2 \delta_0} | Re(w_{P(1)}(\lambda_j) w_{P(2)}(-\lambda_j)  |. 
\end{align*}

Thus \eqref{c1} holds if the following conditions are satisfied: 
\begin{align}
     (\beta_{1,0} - \hat{\beta}_1)^2  m^{-1/2}\sum_{j = 1}^m |v_j| \lambda_j^{2 \delta_0}  I_{P(1)}(\lambda_j) = o_p(1) \label{c3} \\
     \hat{\beta}_{2,0}^2   m^{-1/2}\sum_{j = 1}^m |v_j| \lambda_j^{2 \delta_0}  I_{P(2)}(\lambda_j)  = o_p(1)  \label{c4} \\
    2 |(\beta_{1,0} - \hat{\beta}_1)|  m^{-1/2}\sum_{j = 1}^m |v_j| \lambda_j^{2 \delta_0} |Re(w_{u}(\lambda_j) w_{P(1)}(-\lambda_j))|  = o_p(1)  \label{c5} \\
    2 |\hat{\beta}_2|  m^{-1/2}\sum_{j = 1}^m |v_j| \lambda_j^{2 \delta_0} |Re(w_{u}(\lambda_j) w_{P(2)}(-\lambda_j))| = o_p(1)  \label{c6}  \\
      2 |(\beta_{1,0} - \hat{\beta}_1)| |\hat{\beta}_2|    m^{-1/2}\sum_{j = 1}^m |v_j| \lambda_j^{2 \delta_0} | Re(w_{P(1)}(\lambda_j) w_{P(2)}(-\lambda_j)  |  = o_p(1) \label{c7}
\end{align}

By Lemma \ref{consCheby},  $(\hat{\beta}_1-\beta_{1,0}) = O_p(T^{\delta_0 - 1/2})$  and $\hat{\beta}_2 = O_p(T^{\delta_0 - 1/2})$. Therefore, \eqref{c3} and \eqref{c4} follow by the same argument as in the proof of \eqref{b3}. Similarly, \eqref{c5} and \eqref{c6} follow by the same argument as in the proof of \eqref{b4}. Finally, \eqref{c7} follows by the same argument as in the proof of \eqref{a6}. 

\end{proof}

\subsection{Proof of Theorem \ref{l1}}

\begin{proof}

First, consider the case where $\delta_0 \leq 0$. Then $T^{1 - 2\delta_0} \geq T$, and hence $T^{1 - 2\delta_0} A(T) \geq T A(T)$. From Assumption \ref{ass4}, we have that $T A(T) \to \infty$, which implies that $T^{1 - 2\delta_0} A(T) \to \infty$ as well. Therefore, $A(T)$ satisfies Assumption \ref{ass3}, which, according to Theorem \ref{thm3}, implies that $\hat{k}$ is consistent.

Let now $\delta_0 > 0$. In the proof of Theorem \ref{thm3}, it is shown that the information criterion does not select the under-specified model in the limit if $A(T) = o(1)$. Hence, it is sufficient to examine the scenario when $k > k_0$ and compare it to $k = k_0$. 

From \eqref{diffs}, for $k > k_0$, 
\begin{align*}
 \hat{\sigma}^2(k) - \hat{\sigma}^2(k_0) = - T^{-1+2\delta_0} Q_T,
\end{align*}
where $Q_T = \sum_{s = k_0 + 1}^{k}   \left(T^{1/2-\delta_0} \hat{\beta}_{s} \right)^2$ and $Q_T \geq 0$. By Lemma \ref{consCheby}, we have $Q_T = O_p(1)$. Moreover, $P(Q_T > 0) \rightarrow 1$. To see this, note that under the FCLT the vector $T^{1/2-\delta_0} \hat{\beta}(k)$ converges in distribution to a non-degenerate Gaussian vector, say $Z$. Hence, by the continuous mapping theorem, $Q_T \overset{d}{\rightarrow} ||Z||^2$. Since $Z$ is non-degenerate Gaussian, $P(||Z||^2 = 0) = 0$, which implies $P(Q_T > 0) \rightarrow 1$.  

Hence, from \eqref{vc1}, we have
\begin{align*}
T \ln\left(\frac{\hat{\sigma}^2(k)}{\hat{\sigma}^2(k_0)}\right) = -\frac{T^{2\delta_0} Q_T}{\sigma^2_u} + O_p(T^{-1+4\delta_0}) \xrightarrow{p}  -\infty,
\end{align*}
for any $\delta_0 > 0$.

Given that $T A(T) = o(T^{2\delta_0})$ as $T \rightarrow \infty$, the first term in
\begin{align*}
    T\left(IC(k) - IC(k_0)\right) = T\ln\left(\frac{\hat{\sigma}^2(k)}{\hat{\sigma}^2(k_0)}\right) + (k - k_0) T A(T)
\end{align*}
dominates and diverges to negative infinity. This implies that, in the limit, the selected number of polynomial functions exceeds the true number for any $\delta_0 > 0$.

\end{proof}

\subsection{Proof of Theorem \ref{thm3}}

\begin{proof}

It can be readily shown that when $k < k_0$  
\begin{align*}
    \hat{\sigma}^2(k) = \sigma^2_u + F + o_p(1),
\end{align*}
where $\sigma^2_u = E(u^2_t)$ and $F > 0$ is a deterministic constant capturing the contribution of the omitted Chebyshev term(s). The expansion follows by combining: (i) the rate $\hat{\beta}(k)-\beta_0(k) = O_p(T^{\delta_0-1/2})$ from Lemma \ref{consCheby} (ii) the properties of the Chebyshev functions, $T^{-1}\sum_{t = 1}^T P_t(m) P_t(n)$, in \eqref{crossp} (iii) the bound for the cross term $T^{-1}\sum_{t = 1}^T P_t(n) u_t$  in \eqref{crossup} and (iv) and the law of large numbers $T^{-1} \sum_{t = 1}^T u^2_t \rightarrow \sigma^2_u$. The last statement follows because, by Assumption \ref{ass2}, $u_t$ is stationary and ergodic and since $|\delta_0|<1/2$ implies $\sigma^2_u < \infty$. Therefore, Theorem 2.2 in \textcite{hassler2019time} applied with $x_t = u^2_t$ yields $T^{-1} \sum_{t = 1}^T u^2_t \rightarrow \sigma^2_u$ almost surely and hence in probability.

For the correct specified model, i.e.\ $k = k_0$, the same arguments shows that the omitted-component term is absent and therefore
\begin{align*}
    \hat{\sigma}^2(k_0) = \sigma^2_u + o_p(1).
\end{align*}
Under Assumption \ref{ass3}, where $A(T) = o(1)$, the information criteria is therefore not selecting the under-specified model in the limit. Hence, it is sufficient to compare the scenarios $k > k_0$ and $k = k_0$. To that end, consider 
\begin{align*}
    IC(k) - IC(k_0) = \ln\left(\frac{\hat{\sigma}^2(k)}{\hat{\sigma}^2(k_0)}\right) + (k-k_0)A(T).
\end{align*}
with $k > k_0$.
Expressing the first term as
\begin{align*}
     \ln\left(1+\frac{\hat{\sigma}^2(k)-\hat{\sigma}^2(k_0)}{\hat{\sigma}^2(k_0)}\right). 
\end{align*}
Note that for $k > k_0$, the true coefficients satisfy $\beta_{s,0} = 0$ for all $s = k_0 + 1, \ldots, k$. Hence 
\begin{align*}
     \hat{u}_t(k) = u_t + \sum_{n = 0}^{k_0} (\beta_{n,0}-\hat{\beta}_n) P_t(n)-\sum_{s = k_0 + 1}^{k}   \hat{\beta}_{s} P_t(s).
\end{align*}
Since $\hat{u}_t(k_0) = u_t + \sum_{n = 0}^{k_0} (\beta_{n,0}-\hat{\beta}_n) P_t(n)$, it follows that
\begin{align*}
    \hat{u}_t(k) = \hat{u}_t(k_0)-\sum_{s = k_0 + 1}^{k}   \hat{\beta}_{s} P_t(s).
\end{align*}
Therefore
\begin{align*}
   \hat{\sigma}^2(k) = \hat{\sigma}^2(k_0) + \frac{1}{T} \sum_{t = 1}^T \left( \sum_{s = k_0 + 1}^{k}   \hat{\beta}_{s} P_t(s) \right)^2 - 2  \frac{1}{T} \sum_{t = 1}^T \hat{u}_t(k_0) \sum_{s = k_0 + 1}^{k}   \hat{\beta}_{s} P_t(s).
\end{align*}
Using the properties of the Chebyshev polynomials in \eqref{crossp}, we have for $s \geq 1$, $T^{-1} \sum_{t = 1}^T P^2_t(s) = 1$, $T^{-1} \sum_{t = 1}^T P_t(s) P_t(r) = 0$ for $s \neq r$, and $\hat{\beta}_s = \sum_{t = 1}^T  u_t P_t(s)$ for $s \geq k_0+1$. Therefore
\begin{align}
    \hat{\sigma}^2(k) - \hat{\sigma}^2(k_0) &= \sum_{s = k_0 + 1}^{k}   \hat{\beta}^2_{s} - 2  \sum_{s = k_0 + 1}^{k} \hat{\beta}_{s}  \frac{1}{T} \sum_{t = 1}^T  u_t P_t(s) \nonumber\\
    &= -\sum_{s = k_0 + 1}^{k}   \hat{\beta}^2_{s}. \label{diffs}
\end{align}
By Lemma \eqref{consCheby}, $ \hat{\beta}_{s} = O_p(T^{\delta_0-1/2})$ for $s = k_0+1,\ldots,k$.
Therefore,
\begin{align*}
    \hat{\sigma}^2(k) - \hat{\sigma}^2(k_0) = O_p(T^{2\delta_0-1}).
\end{align*}
Since $\hat{\sigma}^2(k_0) = \sigma_u^2 + o_p(1)$, we have 
\begin{align*}
     \frac{\hat{\sigma}^2(k)-\hat{\sigma}^2(k_0)}{\hat{\sigma}^2(k_0)} = O_p(T^{2\delta_0-1}) = o_p(1)
\end{align*}
because $|\delta_0|<1/2$.

Using
\begin{align*}
    \ln(1+x) = x + O(x^2) \label{vc1}
\end{align*}
for $x = o_p(1)$, it follows that
\begin{align}
     \ln\left(\frac{\hat{\sigma}^2(k)}{\hat{\sigma}^2(k_0)}\right) = O_p(T^{2\delta_0-1}). 
\end{align}
In order for $T^{1-2\delta_0} (IC(k) - IC(k_0)) \rightarrow \infty$ it is therefore sufficient to have $T^{1-2\delta_0}A(T) \rightarrow \infty$ as $T \rightarrow \infty$. This implies that $P(IC(k) > IC(k_0))\rightarrow 1$.
\end{proof}

\subsection{Proof of Theorem \ref{thm5}}
\begin{proof}
Consider $k < k_0$. We first show that 
\begin{align*}
      R_{\hat{u}(k)} (\hat{\delta}(k);m) -  R_{u} (\delta_0;m)  > 0. 
\end{align*}
Since $\delta_0 < \hat{\delta}(k) \leq \Delta_2 < 1/2$ with probability tending to 1,
it suffices to show that 
\begin{align*}
      R_{\hat{u}(k)} (\delta;m) -  R_{u} (\delta_0;m)  > 0 
\end{align*}
for $\delta \in (\delta_0, \Delta_2]$.
Rewriting gives
\begin{align*}
      R_{\hat{u}(k)} (\delta;m) -  R_{u} (\delta_0;m) &= \ln\left(  \frac{1}{m} \sum_{j = 1}^m \left(\frac{j}{m}\right)^{2(\delta - \delta_0)} \frac{I_{\hat{u}(k)}(\lambda_j)}{G_0 \lambda_j^{-2\delta_0}} \right) - \ln\left(  \frac{1}{m} \sum_{j = 1}^m \frac{I_{u}(\lambda_j)}{G_0 \lambda_j^{-2\delta_0}} \right)\\
      & \ \ \ - 2 \left( \delta - \delta_0 \right) \frac{1}{m} \sum_{j = 1}^m \ln\left(\frac{j}{m}\right) \\
    &= \ln\left(  \frac{1}{m} \sum_{j = 1}^m \left(\frac{j}{m}\right)^{2(\delta - \delta_0)} \frac{I_{\hat{u}(k)}(\lambda_j)}{G_0 \lambda_j^{-2\delta_0}} \right) - \ln\left(  \frac{1}{m} \sum_{j = 1}^m \frac{I_{u}(\lambda_j)}{G_0 \lambda_j^{-2\delta_0}} \right)\\
      & \ \ \ + 2 \left( \delta - \delta_0 \right) + o(1),
\end{align*}
since $1/m \sum_{j = 1}^m \ln(j/m) = -1 + o(1)$, see \textcite[Proof of Theorem 1]{dalla2006consistent}. Furthermore, by \textcite[Lemma 2]{dalla2006consistent}
\begin{align}
    \frac{1}{m} \sum_{j = 1}^m \left(\frac{j}{m}\right)^{2(\delta - \delta_0)} \frac{I_{u}(\lambda_j)}{G_0 \lambda_j^{-2\delta_0}} \overset{p}{\rightarrow} \left( 1 + 2(\delta-\delta_0) \right)^{-1} \label{stochasticstrenght}
\end{align}
for $\delta-\delta_0 \geq -1/2 + \epsilon$ (any $\epsilon>0$). Hence we may write the decomposition
\begin{align}
      R_{\hat{u}(k)} (\delta;m) -  R_{u} (\delta_0;m) 
    = \ln\left(  \frac{1}{m} \sum_{j = 1}^m \left(\frac{j}{m}\right)^{2(\delta - \delta_0)} \frac{I_{\hat{u}(k)}(\lambda_j)}{G_0 \lambda_j^{-2\delta_0}} \right) + 2 \left( \delta - \delta_0 \right) + o_p(1). \label{decomp1}
\end{align}

For simplicity, we present the argument for $k_0 = 1$ and $k = 0$. The general case follows by the same reasoning. Here, $\hat{u}_t(0) = u_t - (\hat{\beta}_0 - \beta_{0,0})P_t(0)  + \beta_{1,0}P_t(1)$. For $j = 1,\ldots,m$, 
\begin{align}
    I_{\hat{u}(0)}(\lambda_j) &= \left|w_{u}(\lambda_j) + \beta_{1,0} w_{P(1)}(\lambda_j)\right|^2. \label{stochc}
\end{align}
By the reverse triangle inequality and using \eqref{stochasticstrenght},
\begin{align}
     \frac{1}{m} \sum_{j = 1}^m \left(\frac{j}{m}\right)^{2(\delta - \delta_0)} \frac{I_{\hat{u}(0)}(\lambda_j)}{G_0 \lambda_j^{-2\delta_0}} &\geq \left( \sqrt{ \frac{1}{m} \sum_{j = 1}^m \left(\frac{j}{m}\right)^{2(\delta - \delta_0)} \frac{I_{u}(\lambda_j)}{G_0 \lambda_j^{-2\delta_0}}}-|\beta_{1,0}| \sqrt{ \frac{1}{m} \sum_{j = 1}^m \left(\frac{j}{m}\right)^{2(\delta - \delta_0)} \frac{I_{P(1)}(\lambda_j)}{G_0 \lambda_j^{-2\delta_0}}}\right)^2 \\
     &=  \left( \sqrt{ \left( 1 + 2(\delta-\delta_0) \right)^{-1} }-|\beta_{1,0}| \sqrt{ \frac{1}{m} \sum_{j = 1}^m \left(\frac{j}{m}\right)^{2(\delta - \delta_0)} \frac{I_{P(1)}(\lambda_j)}{G_0 \lambda_j^{-2\delta_0}}} + o_p(1)\right)^2 . \label{decomp2}
\end{align}
By Lemma \ref{w_P},
\begin{align}
    \frac{1}{m} \sum_{j = 1}^m \left(\frac{j}{m}\right)^{2(\delta - \delta_0)} \frac{I_{P(1)}(\lambda_j)}{G_0 \lambda_j^{-2\delta_0}} \sim
C T^{1-2\delta_0} m^{-(1+2(\delta-\delta_0))}, \label{detc}
\end{align}
for $\delta<\frac{1}{2}$.

To show $R_{\hat{u}(0)} (\delta;m) -  R_{u} (\delta_0;m) > 0$, we compare the deterministic component in \eqref{detc} with the stochastic component in \eqref{stochasticstrenght}. The quantity 
$T^{1-2\delta_0} m^{-(1+2(\delta-\delta_0))}$ must satisfy one of the following regimes. 
 
If $T^{1-2\delta_0} m^{-(1+2(\delta-\delta_0))} \rightarrow \infty$, then the deterministic component in \eqref{detc} diverges, so the lower bound in \eqref{decomp2} diverges to $+\infty$. Hence the logarithmic term in \eqref{decomp1} diverges to $+\infty$, and therefore 
\begin{align*}
    R_{\hat{u}(0)} (\delta;m) -  R_{u} (\delta_0;m) \rightarrow +\infty.
\end{align*}

 If $T^{1-2\delta_0} m^{-(1+2(\delta-\delta_0))} \rightarrow 0$, then the deterministic contribution in \eqref{decomp2} is asymptotically negligible, and \eqref{decomp1} gives
\begin{align}
    R_{\hat{u}(0)} (\delta;m) -  R_{u} (\delta_0;m) 
    \geq -\ln\left( 1 + 2(\delta-\delta_0) \right) + 2 \left( \delta - \delta_0 \right) + o_p(1) > 0, \label{situation1}
\end{align}
since $-\ln(1+x)+x>0$ for $x >0$.

If $T^{1-2\delta_0} m^{-(1+2(\delta-\delta_0))} \rightarrow C$ for some $C>0$, then the stochastic and deterministic terms in \eqref{decomp2} are of same order. In this case expand
\begin{align}
    \frac{1}{m} \sum_{j = 1}^m \left(\frac{j}{m}\right)^{2(\delta - \delta_0)} \frac{I_{\hat{u}(0)}(\lambda_j)}{G_0 \lambda_j^{-2\delta_0}} &=  \frac{1}{m} \sum_{j = 1}^m \left(\frac{j}{m}\right)^{2(\delta - \delta_0)} \frac{ I_{u}(\lambda_j) }{G_0 \lambda_j^{-2\delta_0}} +  \beta_{1,0}^2 \frac{1}{m} \sum_{j = 1}^m \left(\frac{j}{m}\right)^{2(\delta - \delta_0)} \frac{ I_{P(1)}(\lambda_j)  }{G_0 \lambda_j^{-2\delta_0}} \nonumber \\ 
    & \ \ \ + 2 \beta_{1,0} \frac{1}{m} \sum_{j = 1}^m \left(\frac{j}{m}\right)^{2(\delta - \delta_0)} \frac{ Re(w_{u}(\lambda_j) w_{P(1)}(-\lambda_j) )  }{G_0 \lambda_j^{-2\delta_0}}. \label{a11} 
\end{align}
From Lemma \ref{w_P} and the bound $E\left| \lambda_j^{2\delta_0} I_u(\lambda_j)  \right| <C$, the cross-product term in \eqref{a11} satisfies
\begin{align*}
    \left|\frac{1}{m} \sum_{j = 1}^m \left(\frac{j}{m}\right)^{2(\delta - \delta_0)} \frac{ Re(w_{u}(\lambda_j) w_{P(1)}(-\lambda_j) )  }{G_0 \lambda_j^{-2\delta_0}} \right|  = \begin{cases}
O_p( T^{1/2-\delta_0} m^{-1-2(\delta-\delta_0)}), & 2\delta-\delta_0<0,\\
O_p( T^{1/2-\delta_0} m^{-1+\delta_0} \ln(m)), & 2\delta-\delta_0=0, \\
O_p( T^{1/2-\delta_0} m^{-1+\delta_0} ), & 2\delta-\delta_0>0
\end{cases}
\end{align*}
With $T^{1-2\delta_0} m^{-(1+2(\delta-\delta_0))} \rightarrow C$,
 $|\delta_0| < 1/2$, $\delta < 1/2$, and $\delta > \delta_0$, all three bounds are $o_p(1)$. Hence the cross-product term in \eqref{a11} is $o_p(1)$. Since the second term in \eqref{a11} is nonnegative, we obtain
\begin{align*}
    \frac{1}{m} \sum_{j = 1}^m \left(\frac{j}{m}\right)^{2(\delta - \delta_0)} \frac{I_{\hat{u}(0)}(\lambda_j)}{G_0 \lambda_j^{-2\delta_0}} &\geq  \frac{1}{m} \sum_{j = 1}^m \left(\frac{j}{m}\right)^{2(\delta - \delta_0)} \frac{ I_{u}(\lambda_j) }{G_0 \lambda_j^{-2\delta_0}} +o_p(1) 
\end{align*}
Plugging this into \eqref{decomp1} and using \eqref{stochasticstrenght} again returns to the lower bound in \eqref{situation1}, which is strictly positive. 

Therefore, for $k<k_0$, $R_{\hat{u}(k)} (\delta;m) -  R_{u} (\delta_0;m) > 0$, uniformly over $\delta \in (\delta_0, \Delta_2]$, and hence also 
\begin{align*}
     R_{\hat{u}(k)} (\hat{\delta};m) -  R_{u} (\delta_0;m)  > 0.
\end{align*}

Next we show below in \eqref{q4} that 
\begin{align*}
      R_{\hat{u}(k_0)}(\hat{\delta}(k_0);m) = R_{u}(\delta_0;m) + O_p(m^{-1}).
\end{align*}
Hence, for any fixed $k<k_0$,
\begin{align}
    B(k)-B(k_0)>0, 
\end{align}
and, therefore,
\begin{align*}
\IC(k)-\IC(k_0)=\left(B(k)-B(k_0)\right)+(k-k_0)A(T) > (k-k_0)A(T).
\end{align*}
By Assumption \ref{ass6}, $A(T) = o(1)$. Consequently,
\begin{align*}
P \left(\IC(k)>\IC(k_0)\right) \rightarrow 1,
\end{align*}
for each fixed $k<k_0$, and therefore
\begin{align}
P\left(  \hat{k} < k_0 \right)\rightarrow 0. \label{Bunder}
\end{align}

Now consider $k \geq k_0$. We apply a second-order Taylor expansion of $R_{\hat{u}(k)}(\cdot;m)$ around $\delta_0$:
\begin{align*}
    R_{\hat{u}(k)}(\hat{\delta}(k);m) = R_{\hat{u}(k)}(\delta_0;m) + \frac{d R_{\hat{u}(k)}(\delta_0;m)}{d \delta } (\hat{\delta}(k) - \delta_0) + \frac{1}{2} \frac{d^2R_{\hat{u}(k)}(\bar{\delta});m}{d \delta^2} (\hat{\delta}(k) - \delta_0)^2
\end{align*}
where $\bar{\delta}$ is an intermediate point satisfying $|\bar{\delta} - \delta_0| \leq |\hat{\delta}(k) - \delta_0|$. By Lemma \ref{thm6}, $\hat{\delta}(k) \overset{p}{\rightarrow} \delta_0$, hence $\bar{\delta} \overset{p}{\rightarrow} \delta_0$, and moreover $\hat{\delta}(k) - \delta_0 = O_p(m^{-1/2})$. In addition, using \eqref{expansiond} with \textcite[(84)]{dalla2006consistent},
\begin{align*}
    \sqrt{m} \frac{d R_{\hat{u}(k)}(\delta_0;m)}{d \delta }  &\overset{d}{\rightarrow} \mathcal N(0,4)\\
   \frac{d^2R_{\hat{u}(k)}(\bar{\delta};m)}{d \delta^2}  &\overset{p}{\rightarrow} 4. 
\end{align*}
Consequently,
\begin{align}
    R_{\hat{u}(k)}(\hat{\delta}(k);m) = R_{\hat{u}(k)}(\delta_0;m) + O_p(m^{-1}) \label{over1}.
\end{align}
Furthermore, 
\begin{align}
     \left|R_{\hat{u}(k)}(\delta_0;m) - R_{u}(\delta_0;m)\right| = O_p(m^{-1}). \label{over2}
\end{align}
To see this, note first that \eqref{boundRd_0} with $\tau = 1$ implies
\begin{align*}
    \left|\frac{1}{m} \sum_{j = 1}^m \lambda_j^{2\delta_0} I_{\hat{u}(k)}(\lambda_j) - \frac{1}{m} \sum_{j = 1}^m \lambda_j^{2\delta_0} I_{u}(\lambda_j) \right| = O_p(m^{-1}). \end{align*}
Define 
\begin{align*}
    U_{u(k)}(\delta_0;m) &= \frac{1}{m} \sum_{j = 1}^m \lambda_j^{2\delta_0} I_{\hat{u}(k)}(\lambda_j) \\
   U_{u}(\delta_0;m) &= \frac{1}{m} \sum_{j = 1}^m \lambda_j^{2\delta_0} I_{u}(\lambda_j).
\end{align*}
Then $U_{u(k)}(\delta_0;m) - U_{u}(\delta_0;m) = O_p(m^{-1})$. Moreover, by \eqref{peta} and \eqref{peta0},
\begin{align*}
    U_{u(k)}(\delta_0;m)  &\overset{p}{\rightarrow} G_0 >0. \\
  U_{u}(\delta_0;m)  &\overset{p}{\rightarrow} G_0 >0.
\end{align*}
By the mean value theorem for $\ln(\cdot)$, there exists a random $\zeta$ between $ U_{u(k)}(\delta_0;m)$ and $U_{u}(\delta_0;m)$ such that 
\begin{align*}
    \ln(U_{u(k)}(\delta_0;m)) - \ln(U_{u}(\delta_0;m)) = \frac{U_{u(k)}(\delta_0;m)-U_{u}(\delta_0;m)}{\zeta}.
\end{align*}
Since $U_{u(k)}(\delta_0;m)  \overset{p}{\rightarrow} G_0$ and $U_{u}(\delta_0;m)  \overset{p}{\rightarrow} G_0$, we also have $\zeta \overset{p}{\rightarrow} G_0 > 0$.
Therefore,
\begin{align}
    \left|\ln(U_{u(k)}(\delta_0;m)) - \ln(U_{u}(\delta_0;m))\right| = O_p(m^{-1}).  \label{over3}
\end{align}
Finally, the term $-2\delta m^{-1} \sum_{j = 1}^m \ln(\lambda_j)$ in the objective function $R_u(\delta;m)$ does not depend on $u$, so it cancels when taking the difference $R_{\hat{u}(k)}(\delta_0;m) - R_{u}(\delta_0;m)$. Hence \eqref{over3} implies \eqref{over2}.

Combining \eqref{over1} and \eqref{over2} yields 
\begin{align*}
    R_{\hat{u}(k)}(\hat{\delta}(k);m) = R_{u}(\delta_0;m) + O_p(m^{-1}).
\end{align*}
Therefore, for any fixed $k \geq k_0$,
\begin{align}
B(k)=R_{\hat u(k)}(\hat\delta(k);m)=R_u(\delta_0;m)+O_p(m^{-1}). \label{q4}
\end{align}
Hence, for any fixed $k>k_0$,
\begin{align}
    B(k)-B(k_0)=O_p(m^{-1}), \label{Bover}
\end{align}
and, therefore,
\begin{align*}
\IC(k)-\IC(k_0)=\left(B(k)-B(k_0)\right)+(k-k_0)A(T)=O_p(m^{-1}) + (k-k_0)A(T).
\end{align*}
By Assumption \ref{ass6}, $A(T)m\to\infty$, so $(k-k_0)A(T)$ dominates the $O_p(m^{-1})$ term. Consequently,
\begin{align*}
P \left(\IC(k)>\IC(k_0)\right) \rightarrow 1,
\end{align*}
for each fixed $k>k_0$, and therefore
\begin{align*}
P\left(  \hat{k} > k_0 \right)\rightarrow 0.
\end{align*}
Combing this with \eqref{Bunder} yields
\begin{align*}
P\left(  \hat{k} = k_0 \right)\rightarrow 1.
\end{align*}

\end{proof}

\clearpage
\input{supplement}

\end{document}